\documentclass[sigconf]{acmart}

\AtBeginDocument{%
  \providecommand\BibTeX{{%
    \normalfont B\kern-0.5em{\scshape i\kern-0.25em b}\kern-0.8em\TeX}}}

\copyrightyear{2024}
\acmYear{2024}
\setcopyright{acmlicensed}
\acmConference[WWW '24]{Proceedings of the ACM Web Conference 2024}{May 13--17, 2024}{Singapore, Singapore}
\acmBooktitle{Proceedings of the ACM Web Conference 2024 (WWW '24), May 13--17, 2024, Singapore, Singapore}
\acmDOI{10.1145/3589334.3645640}
\acmISBN{979-8-4007-0171-9/24/05}

\usepackage{multirow}
\usepackage[utf8]{inputenc}
\usepackage[flushleft]{threeparttable}
\usepackage{booktabs}
\usepackage{xcolor, colortbl}
\usepackage{makecell}
\usepackage{listings}
\usepackage{xcolor}
\usepackage{hyperref}
\usepackage{graphicx}
\usepackage{titlesec}
\usepackage{microtype}
\usepackage{subcaption}
\usepackage{enumitem}
\usepackage{stfloats}

\definecolor{codegreen}{rgb}{19,138,7}
\definecolor{codegray}{rgb}{0.5,0.5,0.5}
\definecolor{codepurple}{rgb}{0.58,0,0.82}
\definecolor{backcolour}{rgb}{0.95,0.95,0.92}
\lstdefinestyle{mystyle}{
  backgroundcolor=\color{backcolour}, commentstyle=\color{codegreen},
  keywordstyle=\color{blue},
  numberstyle=\tiny\color{codegray},
  stringstyle=\color{codepurple},
  basicstyle=\ttfamily\footnotesize,
  breakatwhitespace=false,         
  breaklines=true,                 
  captionpos=b,                    
  keepspaces=true,                 
  numbers=left,                    
  numbersep=5pt,                  
  showspaces=false,                
  showstringspaces=false,
  showtabs=false,                  
  tabsize=2
}
\titlespacing\section{0pt}{6pt plus 2pt minus 4pt}{4pt plus 0pt minus 2pt}
\titlespacing\subsection{0pt}{6pt plus 2pt minus 4pt}{4pt plus 0pt minus 2pt}
\titlespacing\subsubsection{0pt}{6pt plus 2pt minus 4pt}{4pt plus 0pt minus 2pt}

\begin{document}

\title{Characterizing Ethereum Upgradable Smart Contracts and Their Security Implications}

\author{Xiaofan Li}
\orcid{0009-0003-5951-1948}
\affiliation{%
  \institution{University of Delaware}
  \streetaddress{316B FinTech Innovation Hub, 591 Collaboration Way}
  \city{Newark}
  \state{Delaware}
  \postcode{19713}
  \country{USA}}
\email{xiaofan@udel.edu}

\author{Jin Yang}
\orcid{0009-0009-1706-1444}
\affiliation{%
  \institution{Syracuse University}
  \streetaddress{4-206 Center for Science and Technology}
  \city{Syracuse}
  \state{New York}
  \postcode{13244}
  \country{USA}}
\email{jyang142@syr.edu}

\author{Jiaqi Chen}
\orcid{0000-0002-6368-6164}
\affiliation{%
  \institution{Syracuse University}
  \streetaddress{4-206 Center for Science and Technology}
  \city{Syracuse}
  \state{New York}
  \postcode{13244}
  \country{USA}}
\email{jchen217@syr.edu}

\author{Yuzhe Tang}
\orcid{0000-0002-8911-106X}
\affiliation{%
  \institution{Syracuse University}
  \streetaddress{4-206 Center for Science and Technology}
  \city{Syracuse}
  \state{New York}
  \postcode{13244}
  \country{USA}}
\email{ytang100@syr.edu}

\author{Xing Gao}
\orcid{0009-0000-2574-029X}
\affiliation{%
  \institution{University of Delaware}
  \streetaddress{316B FinTech Innovation Hub, 591 Collaboration Way}
  \city{Newark}
  \state{Delaware}
  \postcode{19713}
  \country{USA}}
\email{xgao@udel.edu}
  
\renewcommand{\abstractname}{\MakeUppercase{Abstract}}
\begin{abstract}
Upgradeable smart contracts (USCs) have been widely adopted to enable modifying deployed smart contracts. 
While USCs bring great flexibility to developers, improper usage might introduce new security issues, potentially allowing attackers to hijack USCs and their users.
 In this paper, we conduct a large-scale measurement study to characterize USCs and their security implications in the wild.
 We summarize six commonly used USC patterns and develop a tool, USCDetector, to identify USCs without needing source code. 
Particularly, USCDetector collects various information such as bytecode and transaction information to construct upgrade chains for USCs and disclose potentially vulnerable ones.
 We evaluate USCDetector using verified smart contracts (i.e., with source code) as ground truth and show that USCDetector can achieve high accuracy with a precision of 96.26\%. We then use USCDetector to conduct a large-scale study on Ethereum, covering a total of 60,251,064 smart contracts.
USCDetecor constructs 10,218 upgrade chains and discloses multiple real-world USCs with potential security issues.

\end{abstract}


\renewcommand{\keywordsname}{\MakeUppercase{Keywords}}

\maketitle

\section{INTRODUCTION}
\label{sec:introduction}
\newcommand{\ignore}[1]{}

Smart contracts are critical building blocks for decentralized applications (DApps) such as decentralized finance (DeFi)~\cite{BEV} and NFT~\cite{nft_eco}. As of Sep 2023, more than $61$ million smart contracts have been deployed on Ethereum~\cite{bigquery}, the largest blockchain supporting smart contracts. To enforce transparency and trust decentralization, Ethereum and many other contract-supporting blockchains adopt the immutable smart contract design. That is, a smart contract, once deployed, cannot be changed or upgraded by any centralized entities. However, immutability conflicts with various legitimate causes to upgrade a deployed smart contract, such as introducing new functional features or patching security vulnerabilities, leading to inconvenience in practice.
Thus, since 2016, various design patterns for upgradable smart contracts (USCs) have been introduced on Ethereum~\cite{EIP1822, EIP1538, EIP1967, EIP2535}, and widely adopted by DApps~\cite{uniswap,opensea}.
 Also, many third-party libraries, such as OpenZeppelin~\cite{openzeppelin_github}, have been developed to ease and accelerate USC development and deployment.

Despite all these efforts, developing USCs is still challenging and requires developers to be trained with security awareness~\cite{anti-patterns}. 
Otherwise, security vulnerabilities might exist in USCs, allowing attackers to hijack USCs and further affect their users.
 Unfortunately, such vulnerabilities are not rare among USCs.
For example, a widely adopted OpenZeppelin USC template is vulnerable to permanent state impairment that can potentially cause huge financial loss (e.g., over \$50m)~\cite{uups_vulnerability_blog}.
 Attackers have even successfully destroyed a USC and obtained all its ETH~\cite{uups_attacke}.
 With more contracts integrating upgrade features, it becomes more likely that attackers target these upgradeable contracts in the future~\cite{security_guide}.

\ignore{
Smart contracts are fundamental building blocks for decentralized applications such as decentralized finance (DeFi)~\cite{BEV} and NFT~\cite{nft_eco}.
To date, more than 61 million smart contracts have been deployed on Ethereum~\cite{bigquery}.
 Typically, smart contracts are immutable by design once they are deployed on Ethereum. 
While such immutability is necessary for achieving transparency and security, it becomes problematic in many practical situations. 
For example, it makes smart contracts difficult to patch security vulnerabilities or introduce new features. 
To address such situations, upgradeable smart contracts (USCs)~\cite{smartcontract_upgrading} have been introduced in Ethereum since 2016~\cite{Stateofupgrade, gist}. 
Since then, multiple Ethereum Improvement Proposals (EIPs)~\cite{EIP1822, EIP1538, EIP1967, EIP2535} have been presented for implementing suitable upgrade methods.
 Multiple different upgrade approaches have been widely adopted by different smart contracts, such as Uniswap~\cite{uniswap}.
 There are also many third-party libraries~\cite{openzeppelin_github} introduced to ease and accelerate USC development and deployment.

Even with all these efforts, developing USCs is still challenging and requires developers to be trained with security awareness~\cite{anti-patterns}. 
Otherwise, security vulnerabilities might exist in USCs allowing attackers to hijack USCs and further affect their users.
 Unfortunately, such vulnerabilities are not rare among USCs.
For example, a widely adopted OpenZeppelin USC template is vulnerable to permanent state impairment that can potentially cause huge financial loss (e.g., over \$50m)~\cite{uups_vulnerability_blog}.
 Attackers have even successfully destroyed a USC and obtained all its ETH~\cite{uups_attacke}.
 With more contracts integrating upgrade features, it becomes more likely that attackers target these upgradeable contracts in the future~\cite{security_guide}.
}

In this paper, we conduct a large-scale measurement study to characterize USCs and their security implications in the wild.
 We first introduce six commonly used USC patterns and their implementations.
 Specifically, our works cover the straightforward method (e.g., to deploy a new contract and migrate states), the Ethereum built-in method (i.e., Metamorphic contract), and four methods that decouple a contract into two sub-contracts (e.g., one immutable contract and one contract that can be modified).
In addition, we present a series of security risks that could potentially cause serious consequences.
 Some vulnerabilities might enable off-path attackers to completely destroy target USCs (e.g., deny their service) and even hijack existing contracts. 
Other issues might put smart contract users into dangerous situations, such as losing assets or trading deprecated tokens.
 To the best of our knowledge, we are the first to systematically investigate several USC security issues on a large scale.

We develop a tool, USCDetector, to identify USCs and their security issues.
 Unlike previous work relying on source code analysis~\cite{ProxyHunting} and can only detect limited USC types (e.g., proxy-based~\cite{ProxyHunting}), USCDetector collects various information such as bytecode and transaction information, which are available for all contracts, to detect six types of USCs, construct their upgrade chains, and disclose potentially vulnerable ones.
 Thus, USCDetector can cover both unverified (i.e., without source code) and verified smart contracts.
 We evaluate USCDetector using a subset of verified smart contracts, and show that it can achieve high accuracy with an overall 96.26\% precision.

We adopt USCDetector on Ethereum, covering a total of 60,251,064 smart contracts.
USCDetecor constructs 10,218 upgrade chains with 91,959 USCs identified. 
Our results show some interesting observations: while the proxy-based approach is the most popular one, many developers attempt to mix different approaches to implementing USCs, 
which can offer more flexibility and facilitate batch processing. 
Moreover, USCDetector successfully discloses multiple real-world USCs with potential security issues.
 For example, we have identified 15 USCs lacking restrictive checks on the upgradable functions, potentially enabling anyone to hijack them.
 Additionally, we have discovered 118 vulnerable contracts that may completely disable USCs (e.g., become unusable forever).
 We have also identified tokens of many deprecated contracts are still listed in various decentralized exchanges, affecting many unaware users.

\section{BACKGROUND}
\label{sec: background}

\subsection{Ethereum Blockchain}

Blockchain is a public database that records transactions across many nodes in a decentralized network. 
Ethereum is a blockchain embedded with a single canonical computer, referred to as the Ethereum Virtual Machine (EVM)~\cite{evm},  maintaining states that everyone on the Ethereum network agrees on. 
There are two types of accounts on Ethereum: (1) externally owned account (EOA) controlled by anyone with private keys, and (2) contract account controlled by code (i.e., smart contract) ~\cite{evm_accounts}. 
Only an EOA can initiate a transaction, which can be used to transfer ETH or change EVM states.
 Typically, a transaction contains various information, including a unique identifier for the transaction (\textit{transaction hash}), the sender address (\textit{from}), and the receiver address (\textit{to}). 
The {\itshape input} is the data sent along with the transaction. 
Transactions are verified and added to the blockchain based on cryptographic mechanisms, and typically cannot be changed without altering all subsequent blocks. 

\subsection{Ethereum Smart Contracts} 

Ethereum supports the execution of smart contracts, which essentially are computer programs that can automatically execute on EVM.
Every smart contract contains a collection of code (i.e., functions) and data (i.e., states) that resides at a specific address. 
A smart contract can be written by {\itshape Solidity}~\cite{solidity} and then compiled to bytecode, which is the final code deployed on Ethereum and executed by EVM. 
EVM bytecode includes approximately 70 different opcodes for computations and communication with the underlying blockchain. 
Each opcode has two representations, a hex value and a more readable mnemonic. 
 Specifically, the source code of smart contracts may not always be available, but the bytecode is publicly accessible on the blockchain~\cite{getcode}.

Smart contracts can be created or executed with transactions. 
Once a contract is deployed, data can be stored in storage or memory for the function's execution. 
The functions in the deployed contract can get/set data in response to incoming transactions. 
Typically, {\itshape internal} functions can only be accessed within the contract or its derived contracts, while {\itshape external} functions can be called from other contracts/users \cite{smart_contract_anatomy}.
To interact with an external function, the function needs to be identified by the first four bytes of the data sent with a transaction. These first four bytes are called \textit{function selectors}, which are calculated by the function name and type of parameters through Keccak hash (SHA3).

Smart contracts serve as the foundation of decentralized applications (DApps)~\cite{dapps} such as Decentralized Exchanges (DEXs)~\cite{dex}. DEXs are open marketplaces for exchanging tokens, which can represent different assets in Ethereum (e.g., lottery tickets, a fiat currency)~\cite{erc_20}. Typically a smart contract that implements the ERC-20 standard (a standard for fungible tokens)~\cite{eip_20} is referred to as a token contract.

\subsection{Smart Contracts Immutability}

Smart contracts are immutable once they are deployed on Ethereum. One reason is that Ethereum does not provide a built-in way to modify deployed smart contracts until the Constantinople hardfork~\cite{constantinople}. 
Also, transactions for interacting smart contracts are packed into blocks to form the blockchain, which is difficult to modify. 
While the immutability of smart contracts achieves better transparency and security in general, there are certain problematic cases. 
First, it is difficult to fix vulnerable contracts, whose vulnerabilities may be either caused by their own logic problems or introduced by the programming languages (e.g., {\itshape Solidity} ~\cite{solidity_bugs}). 
Second, it is difficult to add new features if the current smart contract can no longer meet the needs of users. 
As a result, several smart contract upgrading methods have been utilized to meet the increasing need for upgrading smart contracts.

\section{UPGRADING PATTERNS AND IMPLEMENTATIONS}
\label{sec: upgrade_pattern}

Smart contract upgrading is to modify the code executed in an address while preserving the contract's states. 
Understanding the implementation details of upgradeable smart contracts (USCs) is necessary to investigate their potential security threats. 
This section demystifies six commonly used upgrading patterns and their implementations.

\begin{figure*}[t]

\hspace{-3mm}
\begin{subfigure}{.23\linewidth}
    \includegraphics[scale=0.11]
    {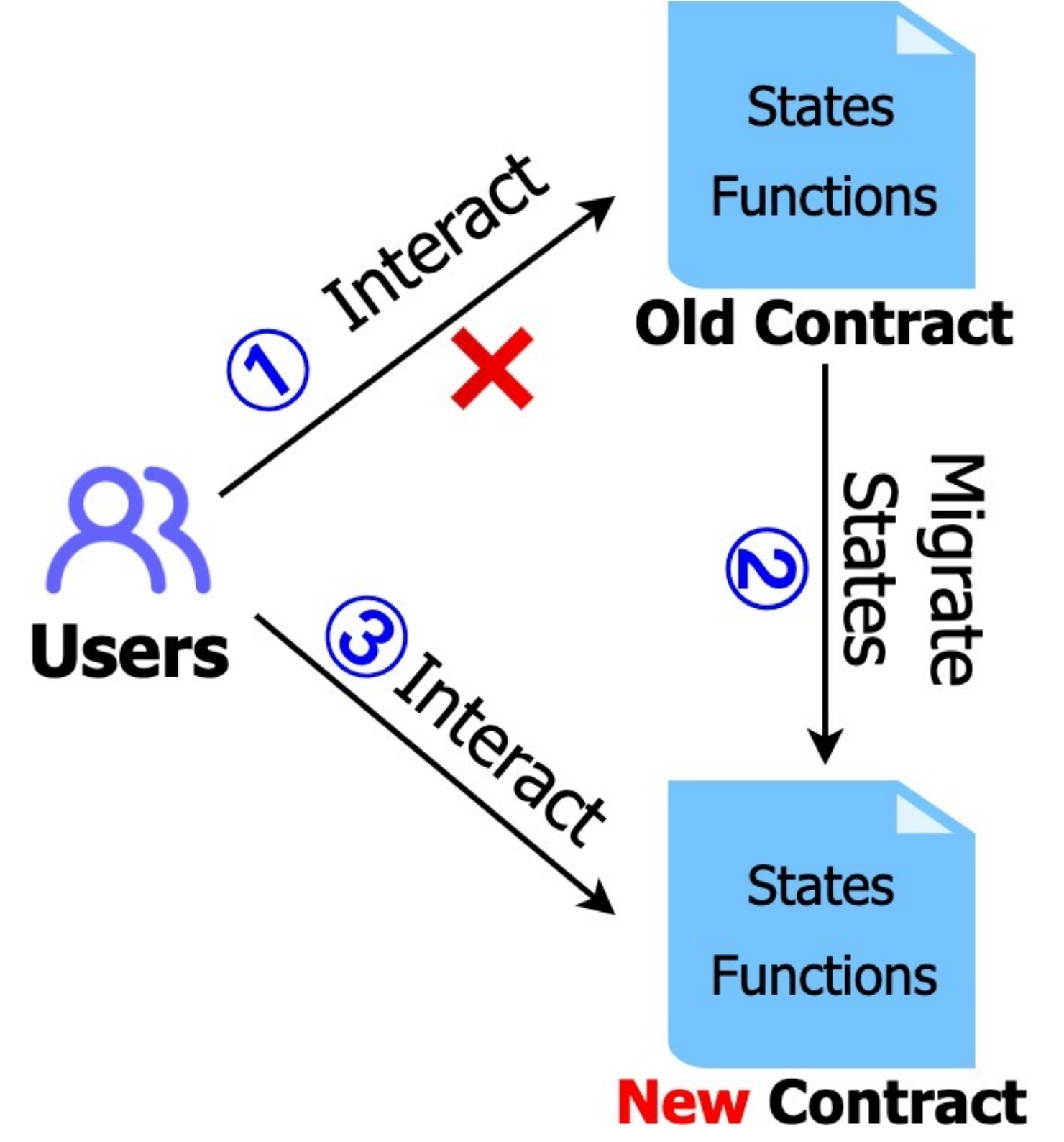}
    \vspace{-2mm}
    \caption{Contract Migration}
    \label{fig: contract_migration}
\end{subfigure}
\hspace{-10mm}
\begin{subfigure}{.23\linewidth}
    \includegraphics[scale=0.11]{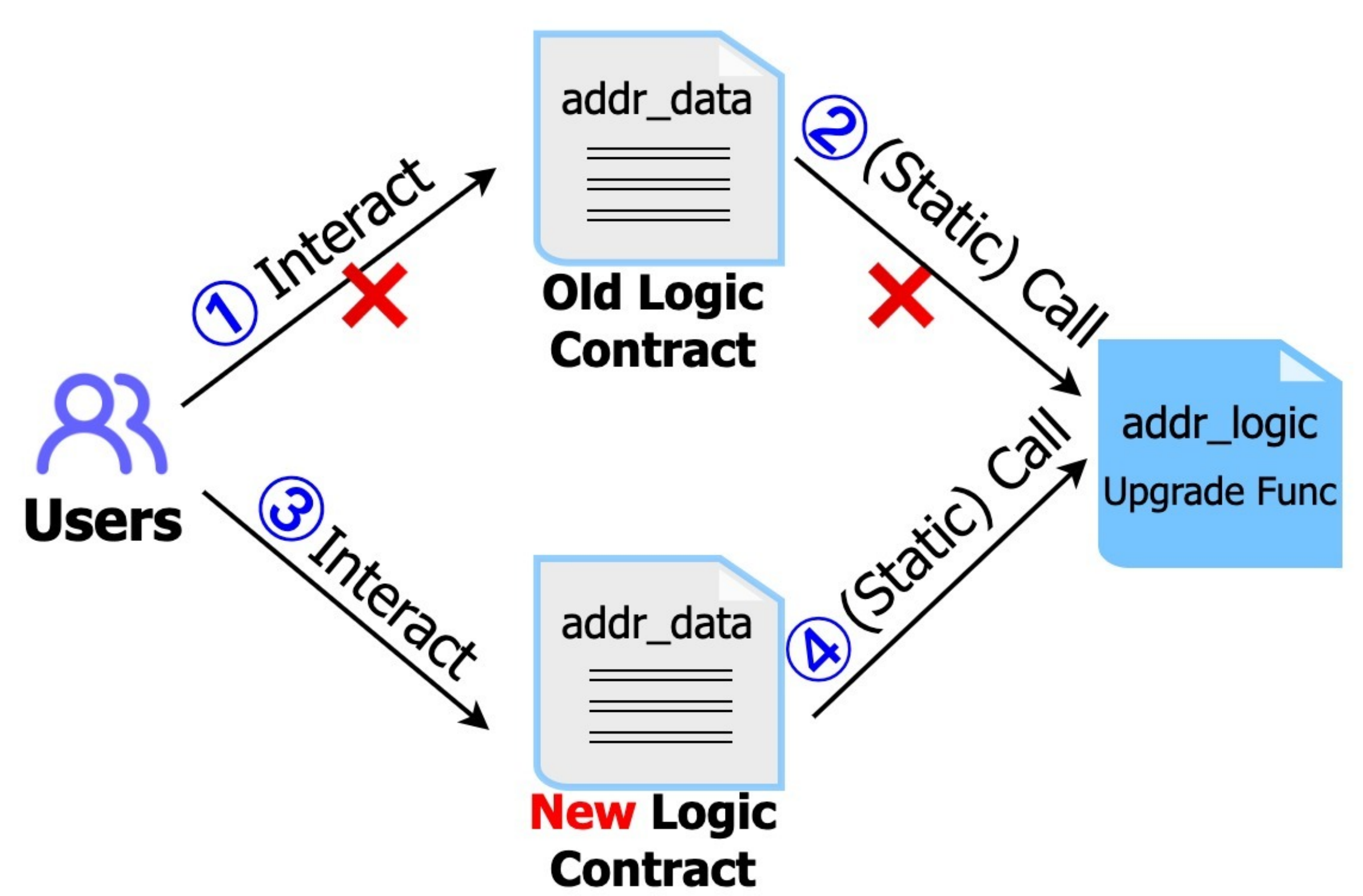}
    \vspace{-5mm}
    \caption{Data Separation}
    \label{fig: data_separation}
\end{subfigure}
\hspace{8mm}
\begin{subfigure}{.23\linewidth}
    \includegraphics[scale=0.11]{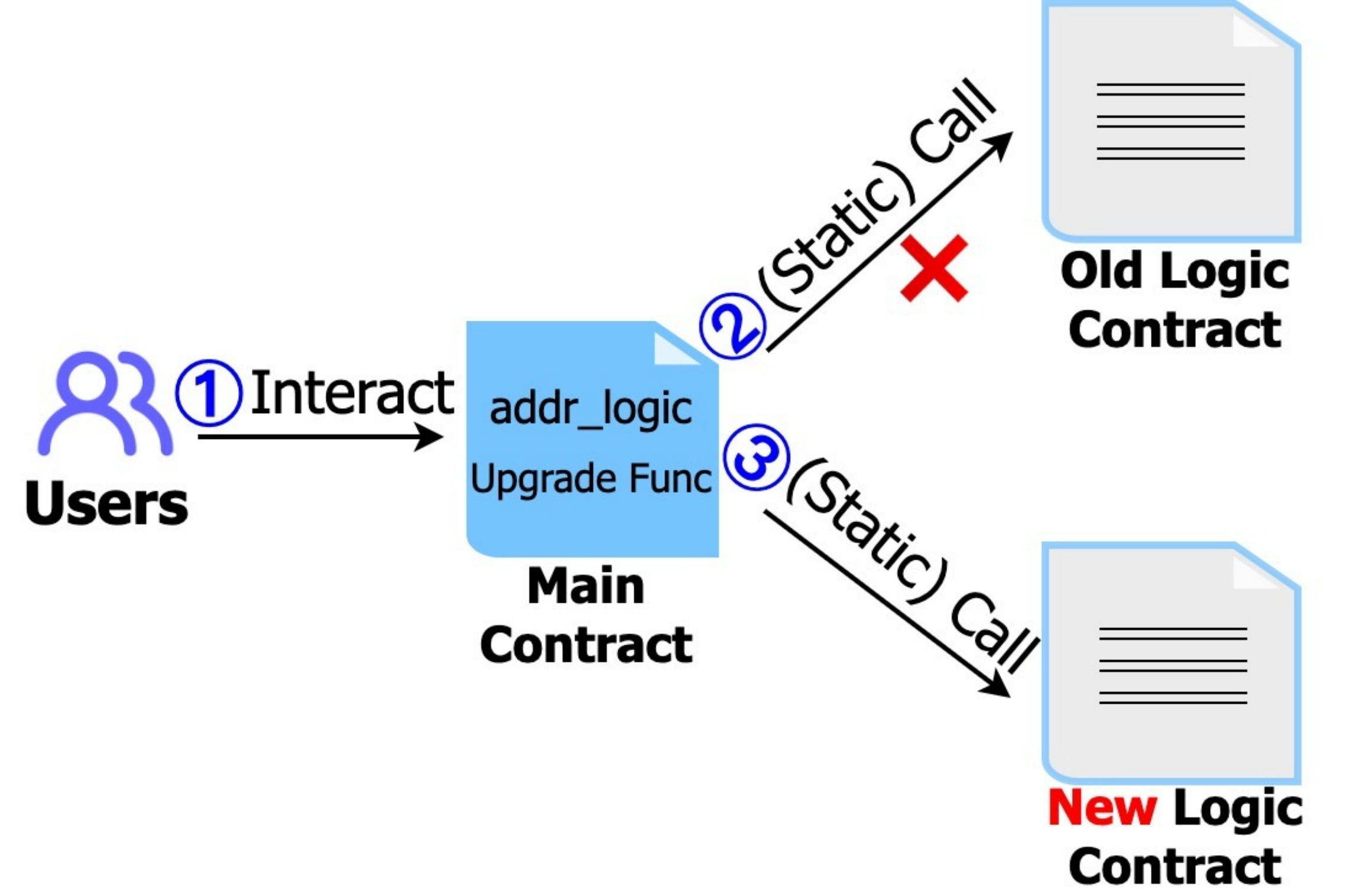}
    \vspace{-5mm}
    \caption{Strategy Pattern}
    \label{fig: strategy_pattern}
\end{subfigure}
\hspace{5mm}
\begin{subfigure}{.23\linewidth}
    \includegraphics[scale=0.11]{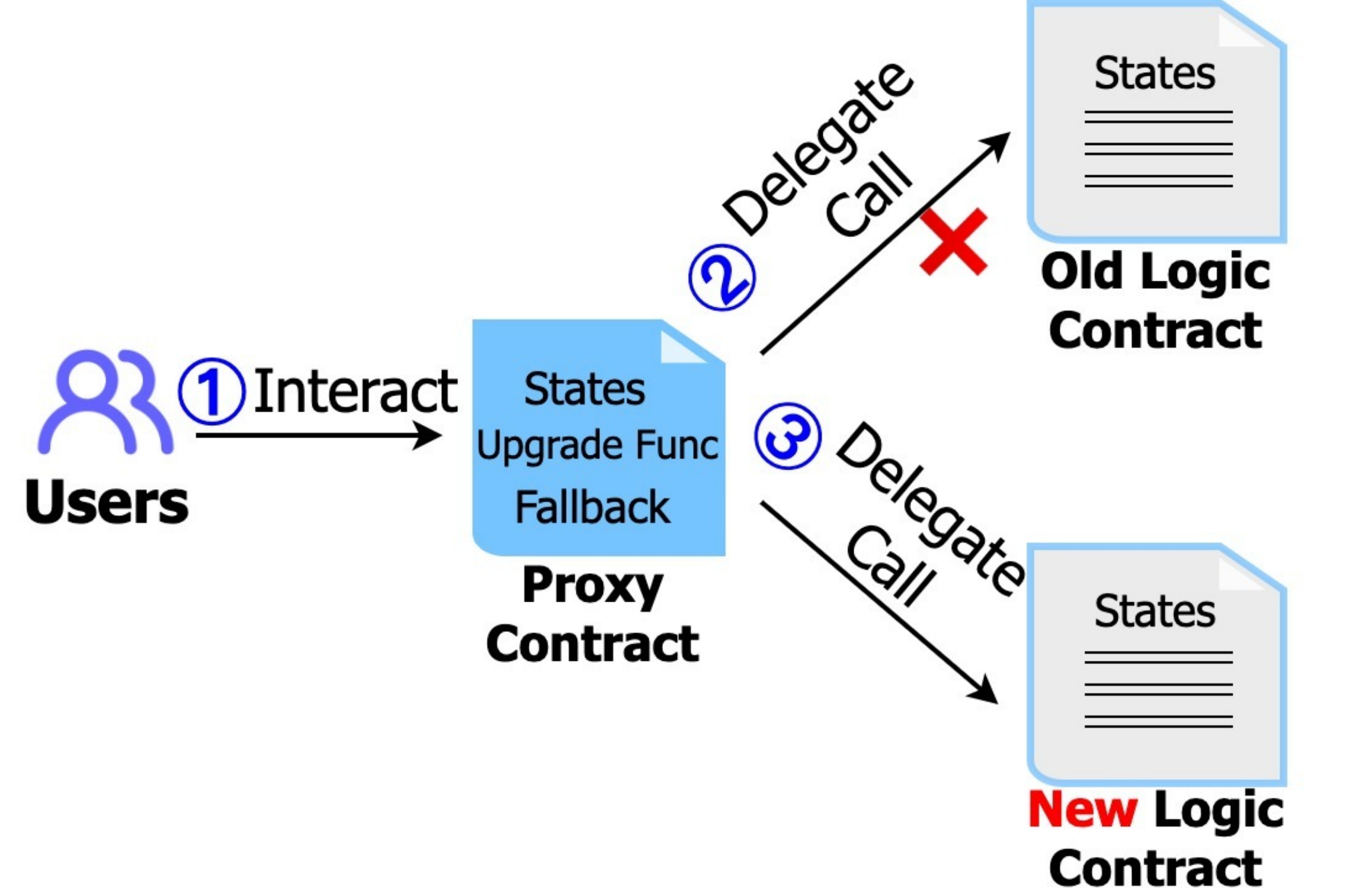}
    \vspace{-5mm}
    \caption{Proxy Pattern}
    \label{fig: proxy_pattern}
\end{subfigure}
\vspace{-3mm}
\caption{USC Patterns.}
\label{fig: pattern_examples}
\vspace{-3mm}
\end{figure*}

\subsection{Contract Migration}
\label{subsec:migration}

The idea of contract migration is to deploy a new smart contract with modified code (i.e., new version), which has an empty state, then migrate all states (e.g., data) from the old (i.e., old version) to the new one. 
Meanwhile, since the new version has a new address, other smart contracts interacting with the old version must also update to the new address, as presented in Figure~\ref{fig: contract_migration}. 
From smart contract users' perspectives, after the migration, the new version contains all users' states (e.g., balances and addresses). Thus, users also need to switch to using the new address. 
For example, if the upgraded contract is a token contract, the old version should be discarded on the exchanges (e.g., Uniswap~\cite{uniswap}), and the new version needs to be listed.

Typically, the deployer needs to make an official announcement claiming that the old version has been migrated to the new one. 
On Etherscan (an Ethereum blockchain explorer)~\cite{etherscan}, only once the deployer provides such information, Etherscan will label the old contract as ``Old Contract''.
Since the new version and the old version are completely independent contracts, it is difficult to detect \textit{contract migration} contracts.
Bandara et al.~\cite{migration_patterns} conducted a study on this pattern, by searching multiple keywords such as ``token'', ``smart contract'', ``migration'' to locate web pages related to blockchain and DApp migration. 
Then they manually analyzed selected web pages to obtain such migration pattern.

\subsection{Data Separation}
\label{subsec:separation}
\textit{Data Separation} is to split a contract into two sub-contracts, with one contract including logic (e.g., code) that can be modified later and the other contract preserves states. 
Users interact with the logic contract, which includes the address of the data contract and uses opcodes \textit{CALL} or \textit{STATICCALL} to interact with it for requiring data.
 The difference is that \textit{STATICCALL} does not allow the called function to change state on EVM.
 Meanwhile, the data contract needs to configure the logic contract's address in its own state. 
Each time a smart contract wants to modify states, the data contract checks whether the caller contract matches its preserved logic address to protect states from being tampered with.

To upgrade, developers simply need to deploy a new logic contract, and then update the address in the data contract's storage. 
As all states are stored in the data contract, there is no need to do extra operations (e.g., state migration). Since users interact with the logic contract, after the upgrade, users also need to switch to the new logic contract (Figure \ref{fig: data_separation}).

\subsection{Strategy Pattern}
\label{subsec:strategy}

\textit{Strategy Pattern} also divides the original smart contract into separate ones: main contract and satellite contracts (i.e., logic contracts). 
The main contract includes both the core business logic and states, as well as the address of the logic contract. 
Similar to data separation, the main contract also interacts with the logic contract to execute certain functions using opcodes \textit{CALL} or \textit{STATICCALL}~\cite{smartcontract_upgrading}.

To upgrade, developers can deploy a new logic contract, and then update the new logic contract's address in the main contract. Since the main contract preserves all states, there is no need to migrate states during the upgrade process (Figure \ref{fig: strategy_pattern}).
 Compared with data separation, the advantage is that users and other smart contracts remain interacting with the main contract and do not need to switch addresses.

\subsection{Proxy Pattern}
\label{subsec:proxy}

Similar to data separation, \textit{proxy pattern} keeps business logic and data in separate contracts (Figure~\ref{fig: proxy_pattern}). 
However, in the proxy pattern, users interact with the storage contract (i.e., \textit{proxy}), which preserves states including the address of the logic contract.
 The proxy contract delegates function calls to the logic contract using the opcode \textit{DELEGATECALL}, which allows the proxy to call the logic contract, while the actual code execution happens in the context of the proxy. 
This means the proxy reads and writes to its own storage.
 It executes logic (e.g., functions) stored at the logic contract similar to calling internal functions~\cite{smartcontract_upgrading}.

In Solidity, a fallback function is executed if the called function does not match any functions in the proxy. 
The proxy can rewrite a custom fallback function that uses \textit{DELEGATECALL} to delegate all unsupported function calls to the logic contract. 
While the proxy is immutable, the address of the logic contract can be replaced with the address of a new logic contract (e.g.,  upgrading~\cite{smartcontract_upgrading}).  
Since the proxy reads and writes to its storage using the logic stored in the logic contract, the function for updating the logic contract's address can be placed either in the proxy or logic contract. 
In particular, if the function for updating the logic contract's address is in the logic contract, this pattern is called \textit{universal upgradeable proxy standard} (UUPS)~\cite{EIP1822, Stateofupgrade}.

To perform an upgrade, developers can deploy a new logic contract with modified code, and then update the new logic contract's address in the proxy contract. Since the proxy contract preserves all states, there is no need to migrate states during the upgrade process.

\noindent\textbf{Third-party Templates.} There are many third-party templates for implementing proxy-based USCs. 
For instance, OpenZeppelin~\cite{openzeppelin_github} is a popular open-source framework providing upgradeable contract templates for developers, such as \textit{UUPSUpgeadeable}~\cite{uupsupgrabeable_sol} and \textit{OwnableUpgradeable}~\cite{ownableupgradeable_sol}.

\subsection{Mix Pattern}
\label{subsec:mix}
Developers can mix the features of both strategy pattern and proxy pattern. 
It can directly call certain functions from the logic contract (via \textit{CALL} or \textit{STATICCALL}), and also delegate calls to the logic contracts (via \textit{DELEGATECALL}).

\subsection{Metamorphic Contract}
\label{subsec:meta}

On 2/28/2019, Ethereum performed Constantinople hardfork~\cite{constantinople} starting from block 7,280,000, and introduced a new opcode {\itshape CREATE2}~\cite{eip1014}, which can be used for upgrading. 
It enables developers to deploy different codes to the same address. 
Developers need to utilize the \textit{SELFDESTRUCT} opcode to wipe out the code and states of that address, then use {\itshape CREATE2} to redeploy code. 
In this way, the bytecode can be changed at the same address (e.g., upgrading). 
This contract is referred to as a Metamorphic contract~\cite{Stateofupgrade}.
 However, this pattern has a drawback that it cannot preserve states after upgrading, as it must destroy the old contract first.

\section{POTENTIAL SECURITY ISSUES}
\label{sec: security}

This section presents several potential security issues in existing USCs, which might be exploited by independent attackers (e.g., not the developers of target USCs) to hijack USCs.
Some might become serious bugs affecting USC users.

\subsection{Missing Restrictive Checks} 
USCs must implement restrictive checks on the upgrade functions ensuring that only contract admin can upgrade contracts. 
Otherwise, anyone can initiate contract upgrading and change the existing logic address to an arbitrary address. 
Attackers can even hijack the contract by changing the logic contract to one controlled by them.

For Metamorphic-based USCs, the first step is to call a function to destroy the Metamorphic contract by using the opcode {\itshape SELFDESTRUCT}.
 Similarly, if this function has no restrictive checks, anyone can potentially destroy this contract.
 Furthermore, once the old contract is destroyed, attackers can hijack the contract by redeploying a new bytecode (e.g., following the Metamorphic-based approach).

\subsection{Insufficient Restrictive Implementation} 
For a proxy-based USC, while users should interact with the proxy, malicious users can send transactions to the logic contract directly. 
In general, this does not pose a threat, since the state of the logic contract does not affect the proxy. 
However, it will become a serious issue, if a malicious user becomes the owner of the logic contract (by initializing it), and then destroys it by calling a function containing \textit{SELFDESTRUCT}. 
The proxy will delegate all calls to a self-destroyed logic contract, causing a denial of service (DoS).
 Particularly, this issue happens if the deployer has only initialized the proxy, but ignored to initialize the logic contract. 
The proxy contract utilizes \textit{DELEGATECALL} to call the initialize function of the logic contract, which runs in the context of the proxy contract, thus leaving the logic contract uninitialized.

There are two cases to destroy the logic contract after being the owner of the logic contract. For case I, the logic contract includes a function containing \textit{SELFDESTRUCT}. A malicious user can simply call this function to destroy the logic contract. 
For case II, the logic contract has a function containing \textit{DELEGATECALL}. Then it can delegate a call to a predefined function containing \textit{SELFDESTRUCT}, which can be called to destroy the logic contract. 
Particularly, case II mainly exists in UUPS and can cause devastating consequences.
 For example, the UUPS template provided by OpenZeppelin~\cite{openzeppelin_github} includes a function \texttt{upgradeToAndCall()}, which introduces such a problem~\cite{uups_vulnerability_blog}).
 For UUPS-based USCs, if the logic contract is destroyed, the USC becomes unavailable forever, as the upgrade function lies in the logic contract.

\subsection{Missing Checks on Logic Address}
The upgrade function needs to conduct necessary logic checks on the contract address of the target logic contract.
 Otherwise, an upgrade may replace the existing logic contract with an arbitrary address, potentially causing irreversible consequences. 
For example, if an upgrade sets the new logic address to an EOA address while the upgrade function is in the logic contract (e.g., UUPS), it can make the \textit{USC} completely unusable. 
We enumerate several issues that are likely to occur with logic addresses as follows.

\vspace{-1mm}
\begin{itemize}[leftmargin=10pt]
\item {\textbf{External owned address.}} The new address is an EOA. When a proxy delegates a call to an EOA, no function will be executed since EOA has no state or code. Thus, this call will always return success, essentially causing a DoS. 
\vspace{-0.5mm}
\item {\textbf{Empty contract.}} The new address is a contract but without any states or functions. Similarly, it causes DoS issues.
\item {\textbf{Zero address.}} The new address is zero (DoS threat). 
\item {\textbf{Same address.}} The same upgrade transaction was submitted multiple times, resulting in duplicate upgrades. Such upgrades might cause a waste of gas.
\item {\textbf{Non-upgradeable logic address in UUPS.}} The new address is zero, or an EOA, or does not contain a function setter. It will make USCs unavailable forever.
\end{itemize}
\vspace{-1.5mm}

\subsection{Contract Version Issues} 
For \textit{contract migration} and \textit{data separation}, the old (logic) contract is replaced by a new contract, which interacts with users. 
Thus, after migration, the old contract should be deprecated.
 For example, Etherscan can publish an announcement indicating that the old version is no longer in use.

\noindent{\textbf{Old contract still in use after migration.}} 
This problem is that, if the old version is not self-destructed, users may not know the upgrade and still interact with the old contract, which might cause severe security issues.
 For example, if the old contract has security vulnerabilities, users' assets in this contract might be in danger.

\noindent{\textbf{Token contracts on DEXs.}}
If the newly deployed contract is a token-based contract, it is important to collaborate with DEXs~\cite{dex} to ensure that the new contract will be listed and the previous one will be discarded. Otherwise, DEXs might still list the old tokens, and future exchange users will swap (e.g., trade) with the old tokens.

\section{METHODOLOGY}
\label{sec: methodology}

We design an analysis tool, USCDetector, to characterize USCs and investigate their potential vulnerabilities in the existing Ethereum blockchain. 
Unlike previous work~\cite{ProxyHunting} that only detect proxy-based USCs based on source code analysis, USCDetector identifies multiple types of USCs based on bytecode, which are always publicly accessible, and thus enable us to characterize both verified and unverified USCs in the wild. 
The high-level idea is that developers attempt to use common keywords (e.g., upgrade, update) in their upgrading functions~\cite{ProxyHunting}, which are available in bytecode. 
With the help of other collectable information (e.g., transactions), we can identify USCs with high accuracy.  

\begin{figure}[t]
  \centering
  \includegraphics[width=0.5\textwidth]{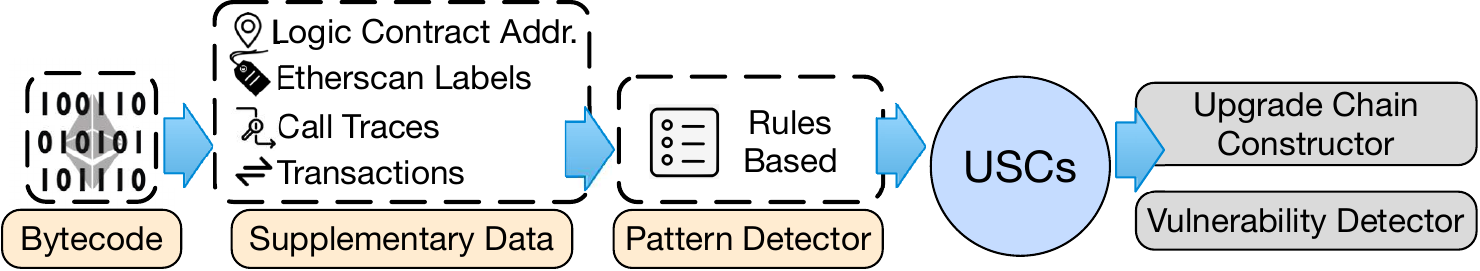}
  \vspace{-6mm}
  \caption{Overview of USCDetector.}
  \label{fig:tool}
    \vspace{-2mm}
\end{figure}

The overall workflow is illustrated in Figure~\ref{fig:tool}.
USCDetector first disassembles smart contract bytecode into opcodes and extracts function selectors.
Then it filters a list of USC candidates based on a set of pre-defined common upgrade function keywords.
 Additionally, USCDetector collects various supplementary information such as logic contract addresses and transaction information.
 Finally, based on all collected information, a rule-based pattern detector identifies USCs and their upgrading patterns.

\subsection{Bytecode Collector} 
\label{subsec:bytecode}
The bytecode collector first uses an Ethereum RPC (Remote Procedure Call) service~\cite{rpc_list} to collect smart contract bytecode.
Then we disassemble bytecode into opcodes using an npm package {\itshape truffle-code-utils}~\cite{truffle-code-utils}, and further utilize  {\itshape abi-decode-functions}~\cite{abi-decode-functions} to extract functions selectors.
Particularly, a smart contract's bytecode first compares the function selector in the transaction's input to all function selectors in the smart contract, and then jumps to the matching function for execution. 
{\itshape abi-decode-functions} extracts function selectors based on such patterns (i.e., match and jump) using pre-defined templates. 
However, we find that it only covers a subset of templates so some function selectors might be missed. 
We thus extend it to include more common patterns.

The above method can extract most function selectors from the immutable contract of USCs (e.g., proxy contract in the proxy pattern, main contract in the strategy pattern, and data contract in the data separation).
However, it cannot extract function selectors of the logic contract in the  \textit{strategy pattern}. 
Particularly, the main contract can use the \textit{PUSH4} opcode to push a 4-byte function selector onto the stack, and then use the opcode \textit{CALL} to call that function whose logic is implemented in the logic contract. 
In other words, the function selector included in the main contract actually represents a function in the logic contract. 
We also extend the {\itshape abi-decode-functions} tool to handle such cases.

Finally, the bytecode collector also extracts various opcodes (e.g., \textit{SELFDESTRUCT}) and other function related information (e.g., fallback function) for pattern detection (Section~\ref{subsec:patterndetector}). 
Table~\ref{ta:opcodes} in Appendix~\ref{app2} lists the detailed description.

\subsection{Common Upgrade Function Keywords.}
\label{subsec:common}
Our idea is to construct and apply a set of keywords that are commonly used in upgrade functions to filter function selectors. Obviously, the quality of such a  dataset is critical for USC detection accuracy. 
We first analyze the Ethereum mainnet dataset~\cite{ProxyHunting}, which contains 2,295 unique (3,822 in total) proxy-based USCs.
From a total of 9,842 upgrade functions, we extract 111 unique upgrade functions, and find most of them can be divided into groups containing five keywords: {\itshape set} (570), {\itshape upgrade} (8,548), {\itshape update} (229), {\itshape change} (88), and {\itshape replace} (22). 
We then query these keywords on the {\itshape Ethereum Signature Database}~\cite{4bytes}, which contains 4-byte signatures of functions in EVM. 
The Ethereum signature database returns all functions including selected keywords and their corresponding 4-byte signatures.
 However, functions containing keywords cannot ensure they are upgrade functions. 
For example, \texttt{setUserData(address,uint256,uint256)} is clearly not used for upgrading. 
Thus, we only keep the functions containing meaningful and related words, such as {\itshape contract}, {\itshape implementation}, and {\itshape logic}, etc.
 We also include some function names that we manually
 collect online, such as {\itshape enableModule} from {\itshape Gnosis Safe Contracts}~\cite{gnosis_modulemanager}.

\subsection{Supplementary Data Collector.} 
We further collect various information in addition to bytecode to assist USC detection.

\noindent\textbf{{Logic Contract Collector.}}
In proxy-based USCs, particularly UUPS, the upgrade function exists in the logic contract, instead of the proxy contract. 
We then utilize an RPC ({\itshape evm-proxy-detection}) to collect the logic contract's address of smart contracts that contain {\itshape DELEGATECALL} opcode (e.g., potential proxy-based USCs).
 Such addresses are further fed to the {\itshape Bytecode Collector} to process their bytecode.

\noindent\textbf{{Call Trace Collector.}}
Metamorphic-based USCs need to destroy the old contract (i.e., using the {\itshape SELFDESTRUCT} opcode) and then redeploy new bytecode using the {\itshape CREATE2} opcode to that address.  
Thus, to identify metamorphic-based USCs, we need to collect the call traces of the transaction that creates a new contract, and then detect if the contract is indeed created by {\itshape CREATE2} in that call traces. 
Particularly, we first request Etherscan API~\cite{etherscancontractapi} to obtain the creators and transaction hashes of smart contracts that contain {\itshape SELFDESTRUCT}. 
Then we use transaction hashes to request Openchain API~\cite{openchainapi} to obtain call traces. 
Finally, we parse the opcodes and input them into the detector.

\noindent\textbf{{Etherscan Crawler.}}
We also crawl various information from Etherscan websites for all contracts that are labeled as ``Old Contract''. 
As mentioned in Section~\ref{subsec:migration}, the  ``Old Contract'' is labeled by Etherscan when the deployer provides the addresses of both the old and new contracts and a link to an official announcement regarding the contract migration~\cite{token_migration}.
 The crawler then collects all related information.

\noindent\textbf{{Transaction Analyzer.}}
We further collect all transactions whose \textit{input} contains upgrade functions (which are potential USCs' upgrade transactions).
 Specifically, we collect them by querying Bigquery~\cite{ethereum_in_bigquery} that uses \textit{ethereum-etl}~\cite{ethereum-etl} to extract data from the Ethereum blockchain every day.
 Then we use \textit{ethereum-input-data-decoder}~\cite{ethereum-input-data-decoder} to decode the transactions' \textit{input} and extract function selectors and arguments.

\subsection{Rule-based Pattern Detector}
\label{subsec:patterndetector}

Based on the characteristics of different USC patterns, we develop a rule-based pattern detector to identify them. 
The detailed rules and notations are listed in Table~\ref{ta:rules} in Appendix~\ref{app1}. 
Specifically, the \textit{proxy pattern} must have (1) both \textit{DELEGATECALL} and the \textit{fallback} function exist in the proxy contract; and (2) an upgrade function in either the proxy or logic contract (i.e., UUPS). 
We first use three criteria for detecting \textit{strategy pattern} and \textit{data separation}: (1) the existence of upgrade functions; (2) \textit{CALL} or \textit{STATICCALL} used; and (3) particular external functions.
 We further distinguish these two patterns using transaction information: strategy-based USCs call logic contracts, while data separation USCs are called by logic contracts.
 For \textit{Metamorphic} contracts, we check whether \textit{SELFDESTRUCT} exists and \textit{CREATE2} is in the call trace. 
If contracts meet the rules for both proxy-based and strategy-based, we mark them as mix patterns.
 Finally, it is difficult to identify the \textit{contract migration} pattern at the bytecode level. Therefore, we select all contracts labeled with the ``Old Contract'' label on Etherscan and remove contracts that are identified as other patterns (e.g., data separation).

\begin{table}[t]
\footnotesize
  \caption{USCDetector Precision on Randomly Sampled Data.}
  \vspace{-2mm}
  \label{ta:random_samp}
  \centering
  \begin{threeparttable}
    \begin{tabular}{llll} 
 \toprule
     \textbf{Patterns} & \textbf{TP} & \textbf{FP} & \textbf{Precision} \\ 
     \midrule[0.5pt]
    \cellcolor[HTML]{DFDFDF}Proxy Pattern & \cellcolor[HTML]{DFDFDF}244 & \cellcolor[HTML]{DFDFDF}6& \cellcolor[HTML]{DFDFDF}97.60\% \\ 
    Data Separation &  43 & 2 &  95.55\% \\ 
    Strategy Pattern &  70 & 2 &  97.22\% \\  
    The Rest of Data or Strategy&  308 & 15 & 95.35\% \\  
     
          \cellcolor[HTML]{DFDFDF}Mix Pattern & \cellcolor[HTML]{DFDFDF}50 & \cellcolor[HTML]{DFDFDF}3 & \cellcolor[HTML]{DFDFDF}94.33\% \\ 
     Metamorphic Contract &  7 & 0 &  100\% \\ 
     \cellcolor[HTML]{DFDFDF}Total &  \cellcolor[HTML]{DFDFDF}722 & \cellcolor[HTML]{DFDFDF}28 & \cellcolor[HTML]{DFDFDF}96.26\% \\ 
     \bottomrule
    \end{tabular}
  \vspace{-3mm}
  \end{threeparttable}
  \end{table}

\subsection{Upgrade Chain Constructor}
\label{subsec:chain}

Finally, we construct contract upgrade chains for USCs that have already performed upgrades.
 The upgrade chain for \textit{contract migration} is straightforward: we simply concatenate contracts based on Etherscan labels. 
For other patterns, to perform an upgrade, an EOA must initiate a transaction to the contract that needs an upgrade. We thus rely on collected transaction information to build the chain.
 For example, upgrading metamorphic contracts is to redeploy new bytecode on the same address. We then chain detected metamorphic contracts with the same address. 
With the upgrade chain, we further check multiple security issues (details in Section~\ref{sec: findings}).

\subsection{USCDetector Evaluation}

The Smart Contract Sanctuary project~\cite{smart_contract_sanctuary} is a project including verified Ethereum smart contracts on Etherscan. 
As of March 22, 2023, this dataset contains 320,080 verified smart contracts, with source code available. 
We input this list into USCDetector and have identified 8,653 USCs, with 2,517 proxy-based USCs, 7 Metamorphic contracts, and 568 mix pattern USCs.
 In addition, there are 5,746 USCs using \textit{data separation} or \textit{strategy pattern}.
 Among them, we further utilize transaction information to separate them. 
Since not all USCs have performed upgrades (i.e., have upgrade transactions), we detect 468 \textit{data separation} and 988 \textit{strategy pattern} USCs.

To evaluate the accuracy, we randomly select and manually verify 750 smart contracts including all patterns based on their source code and decompiled code. 
The result of random sampling is listed in Table~\ref{ta:random_samp}.
 Overall, we are able to achieve 96.26\% precision, with only a few false positives. 
We then randomly select 100 smart contracts from those identified as non-USCs by USCDetector. By examining those contracts, we identified one false negative, which is caused by the exclusion of the upgrading function in our \textit{Common Upgrade Function} dataset.
Appendix~\ref{app4} further presents an evaluation of false negatives using the dataset from Proxy Hunting~\cite{ProxyHunting}. 
It is worth noting that our list of upgrade functions is not obtained from this dataset, indicating that our methodology has the potential to accurately detect unknown USCs without their source code.

\section{MEASUREMENT IN THE WILD}
\label{sec: measurement}

We utilize USCDetector to detect smart contracts  collected from Bigquery~\cite{bigquery}, which has exported 60,251,064 smart contracts (date: 6/5/2023) from the Ethereum blockchain. 
We first group smart contracts based on their bytecode, so that smart contracts in each group have identical bytecode. 
We take the smart contract with the earliest creation time from each group as the representative for further analysis.

\begin{figure*}[t]
  \hspace{-6mm}
    \begin{minipage}{0.38\textwidth}
    \footnotesize
        \captionof{table}{USC Breakdown.}
  \vspace{-2mm}
  \label{ta:total_value}
  \centering
    \begin{tabular}{lll@{}@{}l@{}@{}l} 
 \toprule
     \textbf{Patterns} & \textbf{Raw} & \textbf{Number~} $~$ & \textbf{ETH~} & \textbf{Trans.~} \\ 
     \midrule[0.5pt]
    \cellcolor[HTML]{DFDFDF}Proxy Pattern & \cellcolor[HTML]{DFDFDF}1,866,904 $~$  & \cellcolor[HTML]{DFDFDF}43,650 & \cellcolor[HTML]{DFDFDF}736K & \cellcolor[HTML]{DFDFDF}51M \\
    Data Separation & 1,024 & 1,024 & 131 &  1M \\ 
   Strategy Pattern & 2,444 & 2,444 & 280K &  7.7M \\ 
       The Rest of Data or Strategy$~$& 40,549 & 39,340 & 7K & 11.3M \\
     \cellcolor[HTML]{DFDFDF}Mix Pattern & \cellcolor[HTML]{DFDFDF}23,725 & \cellcolor[HTML]{DFDFDF}1,420 & \cellcolor[HTML]{DFDFDF}2.8K & \cellcolor[HTML]{DFDFDF}5.7M \\ 
     Metamorphic Contract $~$ & 3,097 & 3,097 & 68.5 &  3.3M \\ 
    \cellcolor[HTML]{DFDFDF}Contract Migration & \cellcolor[HTML]{DFDFDF}984 & \cellcolor[HTML]{DFDFDF}984 & \cellcolor[HTML]{DFDFDF}410 & \cellcolor[HTML]{DFDFDF}16M \\ 
    Total & 1,938,727 $~$ & 91,959 &1.025M $~$ & 96M \\
    
     \bottomrule
    \end{tabular}
    \end{minipage}
\hspace{8mm}
    \begin{minipage}{0.29\textwidth}
        \includegraphics[width=1\linewidth]{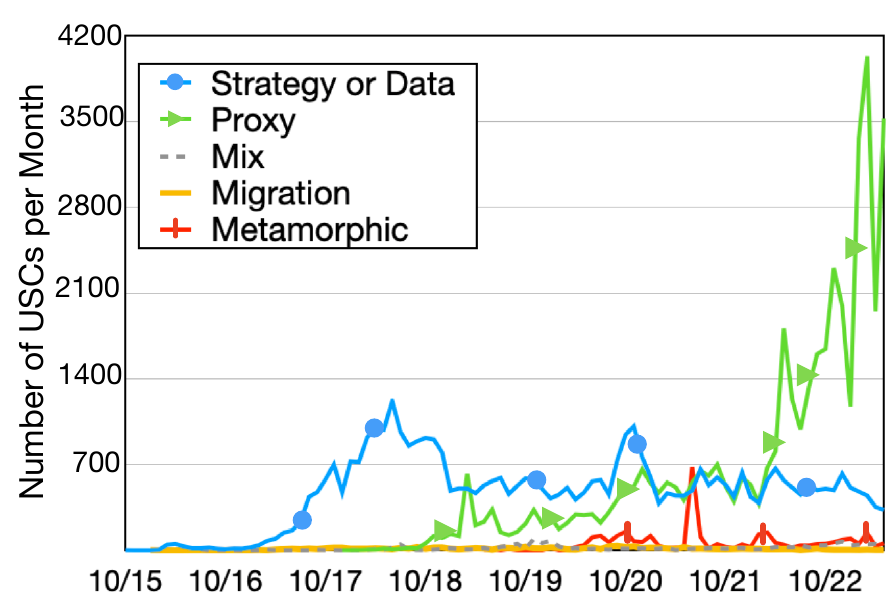}
        \centering
        \vspace*{-8mm}
        \caption{Patterns Over Time.}
        \label{fig: change_number}
    \end{minipage}
        \begin{minipage}{0.28\textwidth}
           \includegraphics[width=0.95\linewidth]{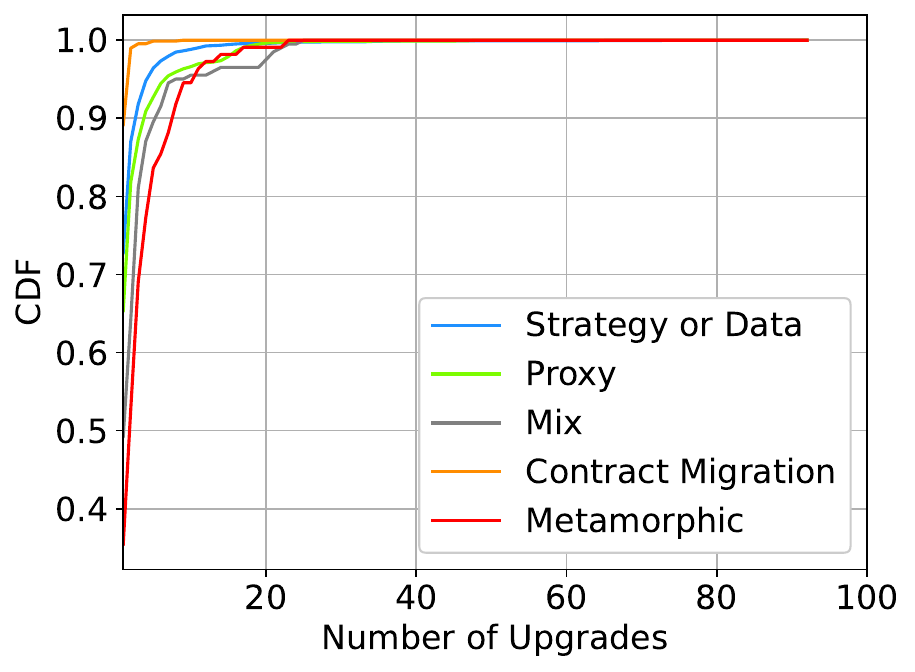}
        \centering
        \vspace*{-4mm}
        \caption{Upgrade Chain CDF.}
        \label{fig:upgrade_chain}
    \end{minipage}
    \hspace{-6mm}
\end{figure*}

In the total of 964,585 groups, we have identified 27,420 groups, with 
1,938,727
individual USCs (column``Raw'' in Table~\ref{ta:total_value}). 
We find that proxy-based pattern dominates existing upgradable methods, with 1,866,904 USCs from 4,964 groups.
 However, most of them are dominated by one group, namely \textit{OwnableDelegateProxy}~\cite{ownabledelegateproxy}, which includes 1,546,462 USCs. 
This group was created by the smart contract {\itshape WyvernProxyRegistry}~\cite{opensearegistry}, which is maintained by OpenSea~\cite{opensea}, a popular NFT market. 
{\itshape WyvernProxyRegistry} creates a proxy contract for each seller on OpenSea (i.e., the seller owns the contract) for executing sellers' actions. 
Obviously, this group of proxies is not for business logic upgrading. In addition to this specific group, there are several other groups adopting a similar approach. In these groups, USCs are created by another smart contract (i.e., a contract factory), instead of an EOA account. The main purpose of a contract factory is to allow DApps (or DeFi) users to generate smart contracts (e.g., create their own tokens ~\cite{contract_factory}). 
In the following analysis (e.g., the numbers presented in Column ``Number'' in Table~\ref{ta:total_value}), we exclude the smart contracts created by a factory smart contract as a practice of deduplication.

\noindent\textbf{Basic Characterization.}
Table~\ref{ta:total_value} shows the detailed breakdown of each pattern.
We also present the aggregated ETH and transaction volumes.
 The most popular is the proxy pattern, with 43,650 contracts containing more than 736k ETH and 51M transactions.
 One possible reason is the wide adoption of third-party templates (e.g., Openzeppelin~\cite{openzeppelin_github}), which provide open source contract libraries for developing smart contracts. 
Figure~\ref{fig: change_number} presents the number of different patterns of USCs over time. The blue line indicates that data separation and strategy pattern were the main upgrade methods until the proxy method (green) was introduced. 
The Metamorphic contract comes after February 2019, and is not popular due to its drawback mentioned in Section~\ref{sec: upgrade_pattern}.

\noindent\textbf{Upgrade Chains.}
In total, we have constructed 4,692 upgrade chains for proxy-based USCs, 4,337 for strategy-based or data separation patterns, 201 for mix pattern,  110 for metamorphic-based USCs, and  878 for contract migration. 
Figure~\ref{fig:upgrade_chain} shows the CDF of upgrade chains. 
Most of them have conducted less than 20 upgrades, while the longest chain contains 92 upgrades.
 Also, 89.1\% of the contract migration approach have only 1 upgrade, which is reasonable as contract migration essentially is to deploy a new contract.

We also find that many upgrades are conducted by different owners (i.e., owner change): there are 185 proxy-based, 97 strategy/data separation, and 52 mix-pattern USCs.

\noindent\textbf{Mix Pattern Demystified.}
We detect 23,725 (Raw) mix-based USCs, which combine features of both strategy and proxy patterns. 
We find that there are different ways to implement the mix pattern.
One popular way (20,591) utilizes an upgradeable intermediate contract. 
The main (proxy) contract can \textit{CALL} the intermediate contract to get the return address for the logic contract.
Then, the main contract can delegate calls to the logic contract using the returned address.
In this way, the main contract does not preserve the actual address of the logic contract. 
When performing an upgrade, developers simply modify the logic contract's address in the intermediate contract. 
It does not require any operations on the main contract for upgrading. 
Thus, it enables uniform upgrades across multiple USCs that share a logic contract.

Another popular approach (704) is similar: the main contract preservers the logic contract's address, and also contains the upgrade function. 
When upgrade, the main contract first \textit{CALL} the logic contract to see if a new logic address is returned. If yes, the main contract updates the new logic address, and then delegate the call to the new one.

\noindent\textbf{Hierarchy Upgrade.} Interestingly, we find some developers utilize strategy-based USCs to further upgrade multiple proxy-based USCs. 
For example, a proxy-based USC already contains an upgrade function (in the proxy contract) that can upgrade its logic contracts. 
Then, the developers utilize a strategy-based main contract to directly \textit{CALL} the upgrade function in the proxy contract to upgrade its logic contract. 
The advantage is that, developers can utilize one main contract to manage multiple proxy contracts.
USCDetector finds 1,349 such strategy-based USCs, and 2,628 proxy-based USCs were upgraded in this way.

\section{SECURITY ISSUE CHARACTERIZATIONS}
\label{sec: findings}

This section presents the security issues related to upgrades. 
We first introduce the methods for identifying them. We also manually verify them based on decompiled code. 

\noindent\textbf{Disclosure.} We have followed the approach of Proxy Hunting~\cite{ProxyHunting} to disclose the vulnerabilities by sending email alerts to contract creators via EthMail~\cite{ethmail}. We have also manually searched their personal social media information to report corresponding issues.

\subsection{Missing Restrictive Checks} 
If a USC misses restrictive checks, an unauthorized user might be able to upgrade this contract.
Based on our collected upgrade chains, we extract USCs that were upgraded by multiple different owners, and further keep USCs if there is an owner who only upgrades the contract once. 
We then manually check their upgrade functions' decompiled code to confirm the missing of restrictive checks.

For metamorphic-based USCs, we check the upgrade chains of metamorphic contracts and locate functions containing \textit{SELFDESTRUCT} from decompiled code. 
Then we check if these functions have or miss restrictive checks.

{\textbf{Results.}} We find that the issue exists in mix pattern (2 USCs), Strategy Pattern (2 USCs), and Metamorphic Contract (11 USCs). 
In total, these USCs have 29 upgrades and attract 59,387 transactions. 
List~\ref{list1} in Appendix~\ref{app2} presents a real-world example derived from DApp LANDProxy~\cite{decentraland}. 
Additionally, we find that many transactions attempt but fail to upgrade USCs: these upgrading transactions are not initialized by the owner(s) performing upgrades.
 Although these transactions fail due to restrictive checks, they indicate that attackers potentially have started to hijack vulnerable USCs.

\subsection{Insufficient Restrictive Implementation.} 

To identify potentially vulnerable logic contracts in proxy-based USCs, we first extract logic contracts that contain \textit{SELFDESTRUCT} for case \uppercase\expandafter{\romannumeral1}, and  UUPS-based logic contracts for case \uppercase\expandafter{\romannumeral2}. 
Then, we query the states of these logic contracts on Oko~\cite{decompiler}, which is an Ethereum explorer listing states of all smart contracts. 
We mark USCs that do not have any state as potentially vulnerable, as it indicates that these logic contracts have not performed any initialization.

Finally, we manually verify whether there is an initialization function that can declare contract ownership in a logic contract from their decompiled code.
Specifically, for case \uppercase\expandafter{\romannumeral1}, if there is such a function in a logic contract, this logic contract is vulnerable as it also includes \textit{SELFDESTRUCT}.
For case \uppercase\expandafter{\romannumeral2}, we consider a logic contract is vulnerable if it contains the function \texttt{upgradeToAndCall} and can be called directly (OpenZeppelin has disabled this function to be called through active proxy after UUPS template version 4.3.2~\cite{uups_vulnerability}).

{\textbf{Results.}} 
For case \uppercase\expandafter{\romannumeral1}, we find 1 vulnerable UUPS-based USC and 51 vulnerable normal proxy-based USCs (i.e., the upgrade function lies in the proxy). 
For case \uppercase\expandafter{\romannumeral2}, we detect 66 vulnerable USCs.
In total, these vulnerable contracts own \$6,350.09 assets (e.g., ETH and tokens), and 12 USCs still have recent transactions on 9/2023.
Note that vulnerabilities in UUPS can completely disable USC (i.e., its proxy contract becomes unusable forever). 
Listings~\ref{list2} and~\ref{list3} in Appendix~\ref{app2} present vulnerable UUPS-based examples of both cases.

\subsection{Missing Checks on Logic Address}
 We utilize the addressing information from upgrade chains to detect logic address issues (e.g., same address, empty contract, EOA, and zero address). 
 Particularly, for empty contracts and EOA, we can not automatically distinguish them. Thus, we manually check the lists with a history of upgrades to non-contracts to classify them.

For detecting non-upgradeable logic addresses in UUPS, we collect all logic addresses from our identified UUPS list, and input them to the {\itshape Pattern Detector} again without enabling rule detection. 
We then filter logic addresses that contain no upgrade function and conduct manual verification.

{\textbf{Results.}} 
We find that these issues are prevalent across all patterns except Metamorphic contracts.  
Table~\ref{ta:logicissue} presents the detailed number.
 The most common problem is successive upgrades to the same address, with a total number of 304.
 These consecutive transactions are often separated by only a few seconds, suggesting that it might be the same upgrade but committed multiple times, causing a waste of gas. 
For example, one proxy-based USC~\cite{proxy_upgrade_four_times} has upgraded four times in a row using the same logic address.
 Also, zero addresses, EOA, and empty contracts are quite common. 
Particularly, as mentioned in Section~\ref{sec: security}, setting the logic address to a non-contract address does not affect other functions in the strategy pattern and data separation. 
Instead, in the proxy and mix patterns, this can cause denial-of-service threats. Figure~\ref{fig: eoa_example} in Appendix~\ref{app5} presents a real-world example that the logic address is an EOA.
In total, these USCs contain over 11.4M transactions with about 226K ETH.

We further explore whether these USCs have corrected their logical addresses later. 
We find that 35 proxies' logic contracts are still EOA accounts, containing more than \$1.5K assets. Particularly,one of them holds an asset of 
\$1.4K.
Even worse, the logic addresses of 15 proxy contracts are actually set to phish accounts (flagged as ``Phish'' by Etherscan) and 12 of them still have transactions. 
 Additionally, one proxy has encountered a malicious transaction from a phishing account, so the creator has set the logic address to zero. 
One proxy's logic address points to a contract that contains no functions or states.

\begin{table}[t]
    \footnotesize
        \captionof{table}{Detected Logic Issues.}
  \vspace{-3mm}
  \hspace*{-2mm}
  \label{ta:logicissue}
  \centering
    \begin{tabular}{l@{}l@{}@{}l@{}@{}l@{}@{}l@{}@{}l} 
 \toprule
     \multirow{2}{*}{\textbf{Patterns}} & \textbf{Same~} & \textbf{Zero~} & \multirow{2}{*}{\textbf{~EOA~}} & \textbf{~Empty~} & \textbf{~Non-Upgrade} \\ 
     & \textbf{Addr.~} &\textbf{Addr.~} & & \textbf{~Contract~} & \textbf{~Addr. in UUPS}  \\
     \midrule[0.5pt]
    \cellcolor[HTML]{DFDFDF}Proxy Pattern &  \cellcolor[HTML]{DFDFDF}160 & \cellcolor[HTML]{DFDFDF}9 & \cellcolor[HTML]{DFDFDF}$~$52 & \cellcolor[HTML]{DFDFDF}$~$3 & \cellcolor[HTML]{DFDFDF}$~$215 \\
    Data Separation &  16 & 1 & $~$3 &  $~$0 & $~$0 \\ 
   Strategy Pattern &  87 & 111 & $~$20 &  $~$0 & $~$0\\ 
       The Rest of Data or Strategy $~$ &  19 & 16 & $~$28 & $~$0 & $~$0\\  
     \cellcolor[HTML]{DFDFDF}Mix Pattern & \cellcolor[HTML]{DFDFDF}22 & \cellcolor[HTML]{DFDFDF}2 & \cellcolor[HTML]{DFDFDF}$~$2 & \cellcolor[HTML]{DFDFDF}$~$0 & \cellcolor[HTML]{DFDFDF}$~$0 \\ 
     Total &  304 & 139 & $~$105 &  $~$3 & $~$215 \\ 
    \bottomrule
    \end{tabular}
    \vspace{-2mm}
    \end{table}

\subsection{Contract Version Issues} 

\noindent{\textbf{Old contract still in use after migration.}} 
From our constructed upgrade chains of \textit{migration-based} USCs, we extract their upgrading reasons from Etherscan announcements.
 As some contracts might use both old and new versions, we only target contracts that explicitly mention that the old version is no longer in use. 
Then, we utilize Etherscan to observe if there are still transactions after the new contracts have been created. 
For \textit{data separation}, we check the usage of old logic contracts after a new logic contract is deployed.

For \textit{contract migration}, we find that 16 old contracts explicitly state that they are no longer in use. However, 10 of them are still interacted with users after publishing their migration announcements, generating a total of 908 transactions. 
 For \textit{data separation}, we detect 21 USCs that have performed upgrades, but users are still interacting with their old logic contracts, generating a total of 253 transactions.
These results demonstrate that \textit{contract migration} and \textit{data separation} are not trivial: users might continue interacting with the old contracts after a successful upgrade.

\noindent{\textbf{Token contracts on DEXs.}}
Token lists~\cite{token_list} is a community-led new standard for creating ERC20 token lists, containing lists from many DEXs like Uniswap and CoinMarketCap~\cite{coinmarketcap}.
From the token list, we pick lists of tokens that have been updated recently, and utilize upgrade chains to check whether old contracts are still listed in these token lists.

 Among 878 constructed upgrade chains (\textit{contract migration}), 
11 have both old and new token contracts listed in at least one DEX; 
45 have their old and new token contracts listed in different DEXs; 
50 have their old token contracts listed in one of the DEXs, but without listing their new token contracts, indicating that old token contracts have not been replaced yet.
 Particularly, one old token contract has 21,275 holders, much more than its new token contract (only 3,794 holders).
 The results indicate that many old tokens are still listed on DEXs, and thus users may swap these old tokens.

\section{RELATED WORKS}
\label{sec: related_works}

\noindent{\textbf{Smart Contract Upgrades.}}
USCs have attracted many research efforts~\cite{Contractpatternssurvey,Specificationislaw,EVMPatch} on understanding their characteristics.
For example, {\itshape Antonino et al.} proposed a framework~\cite{Specificationislaw} to introduce a trusted deployer to vet smart contracts' creation and upgrade. 
{\itshape Rodler et al.} designed EVMPatch~\cite{EVMPatch} to rewrite the bytecode of exploited smart contracts and deploy them as upgradeable proxy contracts. 
Our work focuses on characterizing existing mainstream smart contract upgrading patterns and their security implications. 

One closely related work is Proxy Hunting~\cite{ProxyHunting}, which focuses on security issues of proxy-based USCs from contracts' source code.
Instead, our work covers more USCs patterns (e.g., data separation and strategy pattern) on unverified smart contracts (i.e., without source code). We also investigate several security issues that have not been studied by previous literature. 
Finally, there are two works on detecting metamorphic contracts using opcode {\itshape CREATE2}~\cite{metamorphic_contracts_tool, Notallcode}. 
Our work further utilizes transaction call traces to check whether {\itshape CREATE2} is used to create smart contracts.

\noindent{\textbf{Smart Contract Security Analysis.}}
Extensive research efforts have been conducted on analyzing various security issues (e.g., reentrancy vulnerabilities) of smart contracts~\cite{mythril,maian,verismart,xfuzzer,sailfish,sereum,verx} at both the source code level~\cite{SmartPulse, NPChecker, VETSC, ContractHistory, Clairvoyance, confuzzius, contractfuzzer, osiris, smartchecker,ETHBMC,madmax,callback,reguard,sfuzzer} and bytecode level~\cite{Smarter, EVMPatch, DEFECTCHECKER, Classification, SigRec, smaritian, elysium, smartfield, teether}.
For example, {\itshape Wang et al.} developed NPChecker~\cite{NPChecker} to detect nondeterministic vulnerabilities in smart contracts. 
DefectChecker~\cite{DEFECTCHECKER} is introduced to detect smart contracts' defects from bytecode. 
Different from previous efforts, our work focuses on uncovering existing USCs that have security issues at the bytecode level.

\section{CONCLUSION}
\label{sec: conclusion}
This paper presents a large-scale measurement study on USCs and their potential security issues. 
We have developed USCDetector to identify six types of USC without needing source code. 
Particularly, USCDetector collects various information such as bytecode and transaction information to construct upgrade 
chains for USCs and disclose potentially vulnerable ones.
USCDetector can achieve 96.26\% precision
on our collected ground truth data with verified smart contracts. 
We have used USCDetector to analyze 60,251,064 smart contracts, and discovered multiple real-world USCs that have potential security issues.

\section*{ACKNOWLEDGMENTS}

We would like to thank the anonymous reviewers for their insightful comments. 
Xiaofan Li and Xing Gao are partially supported by National Science Foundation (NSF) grants CNS-2054657 and CNS-2317830.
Jin Yang, Jiaqi Chen, and Yuzhe Tang are partially supported by NSF grants CNS-2139801, CNS-1815814, DGE2104532, and two Ethereum Foundation academic grants.

\renewcommand{\refname}{\MakeUppercase{References}}
\bibliographystyle{ACM-Reference-Format}
\bibliography{ref}


\begin{thebibliography}{90}


\ifx \showCODEN    \undefined \def \showCODEN     #1{\unskip}     \fi
\ifx \showDOI      \undefined \def \showDOI       #1{#1}\fi
\ifx \showISBNx    \undefined \def \showISBNx     #1{\unskip}     \fi
\ifx \showISBNxiii \undefined \def \showISBNxiii  #1{\unskip}     \fi
\ifx \showISSN     \undefined \def \showISSN      #1{\unskip}     \fi
\ifx \showLCCN     \undefined \def \showLCCN      #1{\unskip}     \fi
\ifx \shownote     \undefined \def \shownote      #1{#1}          \fi
\ifx \showarticletitle \undefined \def \showarticletitle #1{#1}   \fi
\ifx \showURL      \undefined \def \showURL       {\relax}        \fi
\providecommand\bibfield[2]{#2}
\providecommand\bibinfo[2]{#2}
\providecommand\natexlab[1]{#1}
\providecommand\showeprint[2][]{arXiv:#2}

\bibitem[uup(2022)]%
        {uups_attacke}
 \bibinfo{year}{2022}\natexlab{}.
\newblock \bibinfo{title}{A Real-World UUPS USC Was Destroyed by Delegating a Call to a Pre-defined Destroy Function}.
\newblock
\newblock
\urldef\tempurl%
\url{https://etherscan.io/address/0xa0e377d9cb4fcc014b634d74de07a428d3896eff}
\showURL{%
\tempurl}


\bibitem[Abi-decoder(2018)]%
        {abi-decode-functions}
\bibfield{author}{\bibinfo{person}{Abi-decoder}.} \bibinfo{year}{2018}\natexlab{}.
\newblock \bibinfo{title}{Abi-decode-functions}.
\newblock
\newblock
\urldef\tempurl%
\url{https://www.npmjs.com/package/abi-decode-functions}
\showURL{%
\tempurl}


\bibitem[Academy(2023)]%
        {contract_factory}
\bibfield{author}{\bibinfo{person}{Solidity Academy}.} \bibinfo{year}{2023}\natexlab{}.
\newblock \bibinfo{title}{Demystifying the Factory Pattern in Solidity: Efficient Contract Deployment with Factory Pattern}.
\newblock
\newblock
\urldef\tempurl%
\url{https://medium.com/@solidity101/demystifying-the-factory-pattern-in-solidity-efficient-contract-deployment-with-factory-pattern-e233ea6d1ec0}
\showURL{%
\tempurl}


\bibitem[Antonino et~al\mbox{.}(2022)]%
        {Specificationislaw}
\bibfield{author}{\bibinfo{person}{Pedro Antonino}, \bibinfo{person}{Juliandson Ferreira}, \bibinfo{person}{Augusto Sampaio}, {and} \bibinfo{person}{AW Roscoe}.} \bibinfo{year}{2022}\natexlab{}.
\newblock \showarticletitle{Specification is Law: Safe Creation and Upgrade of Ethereum Smart Contracts}. In \bibinfo{booktitle}{\emph{International Conference on Software Engineering and Formal Methods}}.
\newblock


\bibitem[Bandara et~al\mbox{.}(2020)]%
        {migration_patterns}
\bibfield{author}{\bibinfo{person}{HMN~Dilum Bandara}, \bibinfo{person}{Xiwei Xu}, {and} \bibinfo{person}{Ingo Weber}.} \bibinfo{year}{2020}\natexlab{}.
\newblock \showarticletitle{Patterns for Blockchain Data Migration}. In \bibinfo{booktitle}{\emph{Proceedings of the European Conference on Pattern Languages of Programs 2020}}.
\newblock


\bibitem[Barros(2019)]%
        {EIP1822}
\bibfield{author}{\bibinfo{person}{Gabriel Barros}.} \bibinfo{year}{2019}\natexlab{}.
\newblock \bibinfo{title}{Universal Upgradeable Proxy Standard (UUPS)}.
\newblock
\newblock
\urldef\tempurl%
\url{https://eips.ethereum.org/EIPS/eip-1822}
\showURL{%
\tempurl}


\bibitem[Blau(2022)]%
        {metamorphic_contracts_tool}
\bibfield{author}{\bibinfo{person}{Michael Blau}.} \bibinfo{year}{2022}\natexlab{}.
\newblock \bibinfo{title}{A Tool for Detecting Metamorphic Smart Contracts}.
\newblock
\newblock
\urldef\tempurl%
\url{https://a16zcrypto.com/posts/article/metamorphic-smart-contract-detector-tool/}
\showURL{%
\tempurl}


\bibitem[Bodell~III et~al\mbox{.}(2023)]%
        {ProxyHunting}
\bibfield{author}{\bibinfo{person}{William~E Bodell~III}, \bibinfo{person}{Sajad Meisami}, {and} \bibinfo{person}{Yue Duan}.} \bibinfo{year}{2023}\natexlab{}.
\newblock \showarticletitle{Proxy Hunting: Understanding and Characterizing Proxy-Based Upgradeable Smart Contracts in Blockchains}. In \bibinfo{booktitle}{\emph{32nd USENIX Security Symposium}}.
\newblock


\bibitem[Bose et~al\mbox{.}(2022)]%
        {sailfish}
\bibfield{author}{\bibinfo{person}{Priyanka Bose}, \bibinfo{person}{Dipanjan Das}, \bibinfo{person}{Yanju Chen}, \bibinfo{person}{Yu Feng}, \bibinfo{person}{Christopher Kruegel}, {and} \bibinfo{person}{Giovanni Vigna}.} \bibinfo{year}{2022}\natexlab{}.
\newblock \showarticletitle{Sailfish: Vetting Smart Contract State-Inconsistency Bugs in Seconds}. In \bibinfo{booktitle}{\emph{2022 IEEE Symposium on Security and Privacy}}.
\newblock


\bibitem[Buterin(2018)]%
        {eip1014}
\bibfield{author}{\bibinfo{person}{Vitalik Buterin}.} \bibinfo{year}{2018}\natexlab{}.
\newblock \bibinfo{title}{Skinny CREATE2}.
\newblock
\newblock
\urldef\tempurl%
\url{https://eips.ethereum.org/EIPS/eip-1014}
\showURL{%
\tempurl}


\bibitem[Center(2023)]%
        {token_migration}
\bibfield{author}{\bibinfo{person}{Etherscan~Information Center}.} \bibinfo{year}{2023}\natexlab{}.
\newblock \bibinfo{title}{Token Migration}.
\newblock
\newblock
\urldef\tempurl%
\url{https://info.etherscan.com/token-migration/}
\showURL{%
\tempurl}


\bibitem[ChainList(2023)]%
        {rpc_list}
\bibfield{author}{\bibinfo{person}{ChainList}.} \bibinfo{year}{2023}\natexlab{}.
\newblock \bibinfo{title}{Ethereum Mainnet RPC and Chain Settings | Chainlist}.
\newblock
\newblock
\urldef\tempurl%
\url{https://chainlist.org/chain/1}
\showURL{%
\tempurl}


\bibitem[Chen(2020)]%
        {ContractHistory}
\bibfield{author}{\bibinfo{person}{Jiachi Chen}.} \bibinfo{year}{2020}\natexlab{}.
\newblock \showarticletitle{Finding Ethereum Smart Contracts Security Issues by Comparing History Versions}. In \bibinfo{booktitle}{\emph{Proceedings of the 35th IEEE/ACM International Conference on Automated Software Engineering}}.
\newblock


\bibitem[Chen et~al\mbox{.}(2021b)]%
        {DEFECTCHECKER}
\bibfield{author}{\bibinfo{person}{Jiachi Chen}, \bibinfo{person}{Xin Xia}, \bibinfo{person}{David Lo}, \bibinfo{person}{John Grundy}, \bibinfo{person}{Xiapu Luo}, {and} \bibinfo{person}{Ting Chen}.} \bibinfo{year}{2021}\natexlab{b}.
\newblock \showarticletitle{Defectchecker: Automated Smart Contract Defect Detection by Analyzing EVM Bytecode}.
\newblock \bibinfo{journal}{\emph{IEEE Transactions on Software Engineering}} (\bibinfo{year}{2021}).
\newblock


\bibitem[Chen et~al\mbox{.}(2021a)]%
        {SigRec}
\bibfield{author}{\bibinfo{person}{Ting Chen}, \bibinfo{person}{Zihao Li}, \bibinfo{person}{Xiapu Luo}, \bibinfo{person}{Xiaofeng Wang}, \bibinfo{person}{Ting Wang}, \bibinfo{person}{Zheyuan He}, \bibinfo{person}{Kezhao Fang}, \bibinfo{person}{Yufei Zhang}, \bibinfo{person}{Hang Zhu}, \bibinfo{person}{Hongwei Li}, {et~al\mbox{.}}} \bibinfo{year}{2021}\natexlab{a}.
\newblock \showarticletitle{SigRec: Automatic Recovery of Function Signatures in Smart Contracts}.
\newblock \bibinfo{journal}{\emph{IEEE Transactions on Software Engineering}} (\bibinfo{year}{2021}).
\newblock


\bibitem[Choi et~al\mbox{.}(2021)]%
        {smaritian}
\bibfield{author}{\bibinfo{person}{Jaeseung Choi}, \bibinfo{person}{Doyeon Kim}, \bibinfo{person}{Soomin Kim}, \bibinfo{person}{Gustavo Grieco}, \bibinfo{person}{Alex Groce}, {and} \bibinfo{person}{Sang~Kil Cha}.} \bibinfo{year}{2021}\natexlab{}.
\newblock \showarticletitle{Smartian: Enhancing Smart Contract Fuzzing with Static and Dynamic Data-Flow Analyses}. In \bibinfo{booktitle}{\emph{2021 36th IEEE/ACM International Conference on Automated Software Engineering}}.
\newblock


\bibitem[Cloud(2018)]%
        {ethereum_in_bigquery}
\bibfield{author}{\bibinfo{person}{Google Cloud}.} \bibinfo{year}{2018}\natexlab{}.
\newblock \bibinfo{title}{Ethereum in BigQuery: A Public Dataset for Smart Contract Analytics}.
\newblock
\newblock
\urldef\tempurl%
\url{https://cloud.google.com/blog/products/data-analytics/ethereum-bigquery-public-dataset-smart-contract-analytics}
\showURL{%
\tempurl}


\bibitem[Cloud(2023)]%
        {bigquery}
\bibfield{author}{\bibinfo{person}{Google Cloud}.} \bibinfo{year}{2023}\natexlab{}.
\newblock \bibinfo{title}{Bigquery}.
\newblock
\newblock
\urldef\tempurl%
\url{https://console.cloud.google.com/bigquery?ws=!1m4!1m3!3m2!1sbigquery-public-data!2scrypto_ethereum}
\showURL{%
\tempurl}


\bibitem[CoinMarketCap(2023)]%
        {coinmarketcap}
\bibfield{author}{\bibinfo{person}{CoinMarketCap}.} \bibinfo{year}{2023}\natexlab{}.
\newblock \bibinfo{title}{Cryptocurrency Prices, Charts and Market Capitalizations}.
\newblock
\newblock
\urldef\tempurl%
\url{https://coinmarketcap.com/}
\showURL{%
\tempurl}


\bibitem[ConsenSys(2023)]%
        {mythril}
\bibfield{author}{\bibinfo{person}{ConsenSys}.} \bibinfo{year}{2023}\natexlab{}.
\newblock \bibinfo{title}{Mythril: Security analysis tool for EVM bytecode}.
\newblock
\newblock
\urldef\tempurl%
\url{https://github.com/ConsenSys/mythril/}
\showURL{%
\tempurl}


\bibitem[CoreLibrary(2021)]%
        {corelibrary}
\bibfield{author}{\bibinfo{person}{CoreLibrary}.} \bibinfo{year}{2021}\natexlab{}.
\newblock \bibinfo{title}{CoreLibrary}.
\newblock
\newblock
\urldef\tempurl%
\url{https://etherscan.io/address/0x57ff2cbf0d1dfd79b497795b2edd3b56f1a30397}
\showURL{%
\tempurl}


\bibitem[Das et~al\mbox{.}(2022)]%
        {nft_eco}
\bibfield{author}{\bibinfo{person}{Dipanjan Das}, \bibinfo{person}{Priyanka Bose}, \bibinfo{person}{Nicola Ruaro}, \bibinfo{person}{Christopher Kruegel}, {and} \bibinfo{person}{Giovanni Vigna}.} \bibinfo{year}{2022}\natexlab{}.
\newblock \showarticletitle{Understanding Security Issues in the NFT Ecosystem}. In \bibinfo{booktitle}{\emph{Proceedings of the 2022 ACM SIGSAC Conference on Computer and Communications Security}}.
\newblock


\bibitem[Database(2023)]%
        {4bytes}
\bibfield{author}{\bibinfo{person}{Ethereum~Signature Database}.} \bibinfo{year}{2023}\natexlab{}.
\newblock \bibinfo{title}{Ethereum Signature Database}.
\newblock
\newblock
\urldef\tempurl%
\url{https://www.4byte.directory/}
\showURL{%
\tempurl}


\bibitem[Decentraland(2023)]%
        {decentraland}
\bibfield{author}{\bibinfo{person}{Decentraland}.} \bibinfo{year}{2023}\natexlab{}.
\newblock \bibinfo{title}{Decentraland}.
\newblock
\newblock
\urldef\tempurl%
\url{https://decentraland.org/}
\showURL{%
\tempurl}


\bibitem[decompiler(2023)]%
        {decompiler}
\bibfield{author}{\bibinfo{person}{Panoramix decompiler}.} \bibinfo{year}{2023}\natexlab{}.
\newblock \bibinfo{title}{Panoramix Decompiler}.
\newblock
\newblock
\urldef\tempurl%
\url{https://oko.palkeo.com/}
\showURL{%
\tempurl}


\bibitem[Docs(2023a)]%
        {smart_contract_anatomy}
\bibfield{author}{\bibinfo{person}{Ethereum Docs}.} \bibinfo{year}{2023}\natexlab{a}.
\newblock \bibinfo{title}{Anatomy of Smart Contracts}.
\newblock
\newblock
\urldef\tempurl%
\url{https://ethereum.org/en/developers/docs/smart-contracts/anatomy/}
\showURL{%
\tempurl}


\bibitem[Docs(2023b)]%
        {erc_20}
\bibfield{author}{\bibinfo{person}{Ethereum Docs}.} \bibinfo{year}{2023}\natexlab{b}.
\newblock \bibinfo{title}{ERC-20 Token Standard}.
\newblock
\newblock
\urldef\tempurl%
\url{https://ethereum.org/en/developers/docs/standards/tokens/erc-20/}
\showURL{%
\tempurl}


\bibitem[Docs(2023c)]%
        {evm_accounts}
\bibfield{author}{\bibinfo{person}{Ethereum Docs}.} \bibinfo{year}{2023}\natexlab{c}.
\newblock \bibinfo{title}{Ethereum Accounts}.
\newblock
\newblock
\urldef\tempurl%
\url{https://ethereum.org/en/developers/docs/accounts/}
\showURL{%
\tempurl}


\bibitem[Docs(2023d)]%
        {evm}
\bibfield{author}{\bibinfo{person}{Ethereum Docs}.} \bibinfo{year}{2023}\natexlab{d}.
\newblock \bibinfo{title}{Ethereum Virtual Machine}.
\newblock
\newblock
\urldef\tempurl%
\url{https://ethereum.org/en/developers/docs/evm/}
\showURL{%
\tempurl}


\bibitem[Docs(2023e)]%
        {dapps}
\bibfield{author}{\bibinfo{person}{Ethereum Docs}.} \bibinfo{year}{2023}\natexlab{e}.
\newblock \bibinfo{title}{Introduction to Dapps}.
\newblock
\newblock
\urldef\tempurl%
\url{https://ethereum.org/en/developers/docs/dapps/}
\showURL{%
\tempurl}


\bibitem[Docs(2023f)]%
        {smartcontract_upgrading}
\bibfield{author}{\bibinfo{person}{Ethereum Docs}.} \bibinfo{year}{2023}\natexlab{f}.
\newblock \bibinfo{title}{Upgrading Smart Contracts}.
\newblock
\newblock
\urldef\tempurl%
\url{https://ethereum.org/en/developers/docs/smart-contracts/upgrading/}
\showURL{%
\tempurl}


\bibitem[Docs(2023g)]%
        {solidity_bugs}
\bibfield{author}{\bibinfo{person}{Solidity Docs}.} \bibinfo{year}{2023}\natexlab{g}.
\newblock \bibinfo{title}{List of Known Bugs}.
\newblock
\newblock
\urldef\tempurl%
\url{https://docs.soliditylang.org/en/latest/bugs.html}
\showURL{%
\tempurl}


\bibitem[Duan et~al\mbox{.}(2022)]%
        {VETSC}
\bibfield{author}{\bibinfo{person}{Yue Duan}, \bibinfo{person}{Xin Zhao}, \bibinfo{person}{Yu Pan}, \bibinfo{person}{Shucheng Li}, \bibinfo{person}{Minghao Li}, \bibinfo{person}{Fengyuan Xu}, {and} \bibinfo{person}{Mu Zhang}.} \bibinfo{year}{2022}\natexlab{}.
\newblock \showarticletitle{Towards Automated Safety Vetting of Smart Contracts in Decentralized Applications}. In \bibinfo{booktitle}{\emph{Proceedings of the 2022 ACM SIGSAC Conference on Computer and Communications Security}}.
\newblock


\bibitem[Ethereum(2023)]%
        {dex}
\bibfield{author}{\bibinfo{person}{Ethereum}.} \bibinfo{year}{2023}\natexlab{}.
\newblock \bibinfo{title}{Decentralized Exchanges (DEXs)}.
\newblock
\newblock
\urldef\tempurl%
\url{https://ethereum.org/en/get-eth/#dex}
\showURL{%
\tempurl}


\bibitem[Etherscan(2023a)]%
        {etherscancontractapi}
\bibfield{author}{\bibinfo{person}{Etherscan}.} \bibinfo{year}{2023}\natexlab{a}.
\newblock \bibinfo{title}{Etherscan Contracts API}.
\newblock
\newblock
\urldef\tempurl%
\url{https://docs.etherscan.io/api-endpoints/contracts}
\showURL{%
\tempurl}


\bibitem[Etherscan(2023b)]%
        {etherscan}
\bibfield{author}{\bibinfo{person}{Etherscan}.} \bibinfo{year}{2023}\natexlab{b}.
\newblock \bibinfo{title}{Etherscan (ETH) Blockchain Explorer}.
\newblock
\newblock
\urldef\tempurl%
\url{https://etherscan.io/}
\showURL{%
\tempurl}


\bibitem[ETHMail(2023)]%
        {ethmail}
\bibfield{author}{\bibinfo{person}{ETHMail}.} \bibinfo{year}{2023}\natexlab{}.
\newblock \bibinfo{title}{Email Services for Ethereum Community}.
\newblock
\newblock
\urldef\tempurl%
\url{https://ethmail.cc/}
\showURL{%
\tempurl}


\bibitem[ETL(2023)]%
        {ethereum-etl}
\bibfield{author}{\bibinfo{person}{Blockchain ETL}.} \bibinfo{year}{2023}\natexlab{}.
\newblock \bibinfo{title}{Ethereum-Etl}.
\newblock
\newblock
\urldef\tempurl%
\url{https://github.com/blockchain-etl/ethereum-etl}
\showURL{%
\tempurl}


\bibitem[Ferreira~Torres et~al\mbox{.}(2022)]%
        {elysium}
\bibfield{author}{\bibinfo{person}{Christof Ferreira~Torres}, \bibinfo{person}{Hugo Jonker}, {and} \bibinfo{person}{Radu State}.} \bibinfo{year}{2022}\natexlab{}.
\newblock \showarticletitle{Elysium: Context-Aware Bytecode-Level Patching to Automatically Heal Vulnerable Smart Contracts}. In \bibinfo{booktitle}{\emph{Proceedings of the 25th International Symposium on Research in Attacks, Intrusions and Defenses}}.
\newblock


\bibitem[Foundation(2023)]%
        {gnosis_modulemanager}
\bibfield{author}{\bibinfo{person}{Safe~Ecosystem Foundation}.} \bibinfo{year}{2023}\natexlab{}.
\newblock \bibinfo{title}{ModuleManager}.
\newblock
\newblock
\urldef\tempurl%
\url{https://github.com/safe-global/safe-contracts/blob/v1.4.1/contracts/base/ModuleManager.sol}
\showURL{%
\tempurl}


\bibitem[Frank et~al\mbox{.}(2020)]%
        {ETHBMC}
\bibfield{author}{\bibinfo{person}{Joel Frank}, \bibinfo{person}{Cornelius Aschermann}, {and} \bibinfo{person}{Thorsten Holz}.} \bibinfo{year}{2020}\natexlab{}.
\newblock \showarticletitle{ETHBMC: A Bounded Model Checker for Smart Contracts}. In \bibinfo{booktitle}{\emph{29th USENIX Security Symposium}}.
\newblock


\bibitem[Fr{\"o}wis and B{\"o}hme(2022)]%
        {Notallcode}
\bibfield{author}{\bibinfo{person}{Michael Fr{\"o}wis} {and} \bibinfo{person}{Rainer B{\"o}hme}.} \bibinfo{year}{2022}\natexlab{}.
\newblock \showarticletitle{Not All Code Are Create2 Equal}. In \bibinfo{booktitle}{\emph{6th Workshop on Trusted Smart Contracts}}.
\newblock


\bibitem[Grech et~al\mbox{.}(2018)]%
        {madmax}
\bibfield{author}{\bibinfo{person}{Neville Grech}, \bibinfo{person}{Michael Kong}, \bibinfo{person}{Anton Jurisevic}, \bibinfo{person}{Lexi Brent}, \bibinfo{person}{Bernhard Scholz}, {and} \bibinfo{person}{Yannis Smaragdakis}.} \bibinfo{year}{2018}\natexlab{}.
\newblock \showarticletitle{MadMax: Surviving Out-of-Gas Conditions in Ethereum Smart Contracts}.
\newblock \bibinfo{journal}{\emph{Proceedings of the ACM on Programming Languages}} (\bibinfo{year}{2018}).
\newblock


\bibitem[Grossman et~al\mbox{.}(2017)]%
        {callback}
\bibfield{author}{\bibinfo{person}{Shelly Grossman}, \bibinfo{person}{Ittai Abraham}, \bibinfo{person}{Guy Golan-Gueta}, \bibinfo{person}{Yan Michalevsky}, \bibinfo{person}{Noam Rinetzky}, \bibinfo{person}{Mooly Sagiv}, {and} \bibinfo{person}{Yoni Zohar}.} \bibinfo{year}{2017}\natexlab{}.
\newblock \showarticletitle{Online Detection of Effectively Callback Free Objects with Applications to Smart Contracts}.
\newblock \bibinfo{journal}{\emph{Proceedings of the ACM on Programming Languages}} (\bibinfo{year}{2017}).
\newblock


\bibitem[History(2019)]%
        {constantinople}
\bibfield{author}{\bibinfo{person}{Ethereum History}.} \bibinfo{year}{2019}\natexlab{}.
\newblock \bibinfo{title}{Constantinople}.
\newblock
\newblock
\urldef\tempurl%
\url{https://ethereum.org/en/history/#constantinople}
\showURL{%
\tempurl}


\bibitem[input-data decoder(2022)]%
        {ethereum-input-data-decoder}
\bibfield{author}{\bibinfo{person}{Ethereum input-data decoder}.} \bibinfo{year}{2022}\natexlab{}.
\newblock \bibinfo{title}{Ethereum-input-data-decoder}.
\newblock
\newblock
\urldef\tempurl%
\url{https://www.npmjs.com/package/ethereum-input-data-decoder}
\showURL{%
\tempurl}


\bibitem[Iosiro(2021)]%
        {uups_vulnerability_blog}
\bibfield{author}{\bibinfo{person}{Iosiro}.} \bibinfo{year}{2021}\natexlab{}.
\newblock \bibinfo{title}{Perma-Brick UUPS Proxies with This One Trick}.
\newblock
\newblock
\urldef\tempurl%
\url{https://www.iosiro.com/blog/openzeppelin-uups-proxy-vulnerability-disclosure}
\showURL{%
\tempurl}


\bibitem[Jiang et~al\mbox{.}(2018)]%
        {contractfuzzer}
\bibfield{author}{\bibinfo{person}{Bo Jiang}, \bibinfo{person}{Ye Liu}, {and} \bibinfo{person}{Wing~Kwong Chan}.} \bibinfo{year}{2018}\natexlab{}.
\newblock \showarticletitle{Contractfuzzer: Fuzzing Smart Contracts for Vulnerability Detection}. In \bibinfo{booktitle}{\emph{Proceedings of the 33rd ACM/IEEE International Conference on Automated Software Engineering}}.
\newblock


\bibitem[Josselinfeist(2018)]%
        {anti-patterns}
\bibfield{author}{\bibinfo{person}{Josselinfeist}.} \bibinfo{year}{2018}\natexlab{}.
\newblock \bibinfo{title}{Contract Upgrade Anti-Patterns}.
\newblock
\newblock
\urldef\tempurl%
\url{https://blog.trailofbits.com/2018/09/05/contract-upgrade-anti-patterns}
\showURL{%
\tempurl}


\bibitem[Krupp and Rossow(2018)]%
        {teether}
\bibfield{author}{\bibinfo{person}{Johannes Krupp} {and} \bibinfo{person}{Christian Rossow}.} \bibinfo{year}{2018}\natexlab{}.
\newblock \showarticletitle{TEEther: Gnawing at Ethereum to Automatically Exploit Smart Contracts}. In \bibinfo{booktitle}{\emph{27th USENIX Security Symposium}}.
\newblock


\bibitem[Liu et~al\mbox{.}(2018)]%
        {reguard}
\bibfield{author}{\bibinfo{person}{Chao Liu}, \bibinfo{person}{Han Liu}, \bibinfo{person}{Zhao Cao}, \bibinfo{person}{Zhong Chen}, \bibinfo{person}{Bangdao Chen}, {and} \bibinfo{person}{Bill Roscoe}.} \bibinfo{year}{2018}\natexlab{}.
\newblock \showarticletitle{Reguard: Finding Reentrancy Bugs in Smart Contracts}. In \bibinfo{booktitle}{\emph{Proceedings of the 40th International Conference on Software Engineering: Companion Proceeedings}}.
\newblock


\bibitem[Luu et~al\mbox{.}(2016)]%
        {Smarter}
\bibfield{author}{\bibinfo{person}{Loi Luu}, \bibinfo{person}{Duc-Hiep Chu}, \bibinfo{person}{Hrishi Olickel}, \bibinfo{person}{Prateek Saxena}, {and} \bibinfo{person}{Aquinas Hobor}.} \bibinfo{year}{2016}\natexlab{}.
\newblock \showarticletitle{Making Smart Contracts Smarter}. In \bibinfo{booktitle}{\emph{Proceedings of the 2016 ACM SIGSAC Conference on Computer and Communications Security}}.
\newblock


\bibitem[Meisami and Bodell~III(2023)]%
        {Contractpatternssurvey}
\bibfield{author}{\bibinfo{person}{Sajad Meisami} {and} \bibinfo{person}{William~Edward Bodell~III}.} \bibinfo{year}{2023}\natexlab{}.
\newblock \showarticletitle{A Comprehensive Survey of Upgradeable Smart Contract Patterns}.
\newblock \bibinfo{journal}{\emph{arXiv preprint arXiv:2304.03405}} (\bibinfo{year}{2023}).
\newblock


\bibitem[Mudge(2018)]%
        {EIP1538}
\bibfield{author}{\bibinfo{person}{Nick Mudge}.} \bibinfo{year}{2018}\natexlab{}.
\newblock \bibinfo{title}{Transparent Contract Standard}.
\newblock
\newblock
\urldef\tempurl%
\url{https://eips.ethereum.org/EIPS/eip-1538}
\showURL{%
\tempurl}


\bibitem[Mudge(2020a)]%
        {EIP2535}
\bibfield{author}{\bibinfo{person}{Nick Mudge}.} \bibinfo{year}{2020}\natexlab{a}.
\newblock \bibinfo{title}{Diamonds, Multi-Facet Proxy}.
\newblock
\newblock
\urldef\tempurl%
\url{https://eips.ethereum.org/EIPS/eip-2535}
\showURL{%
\tempurl}


\bibitem[Mudge(2020b)]%
        {EIP1967}
\bibfield{author}{\bibinfo{person}{Nick Mudge}.} \bibinfo{year}{2020}\natexlab{b}.
\newblock \bibinfo{title}{Proxy Storage Slots}.
\newblock
\newblock
\urldef\tempurl%
\url{https://eips.ethereum.org/EIPS/eip-1967}
\showURL{%
\tempurl}


\bibitem[MVHQ(2022)]%
        {proxy_upgrade_four_times}
\bibfield{author}{\bibinfo{person}{MVHQ}.} \bibinfo{year}{2022}\natexlab{}.
\newblock \bibinfo{title}{A Proxy-Based USC Upgrades Four Times in a Row}.
\newblock
\newblock
\urldef\tempurl%
\url{https://etherscan.io/txs?a=0x2809a8737477a534df65c4b4cae43d0365e52035&p=36}
\showURL{%
\tempurl}


\bibitem[Nguyen et~al\mbox{.}(2020)]%
        {sfuzzer}
\bibfield{author}{\bibinfo{person}{Tai~D Nguyen}, \bibinfo{person}{Long~H Pham}, \bibinfo{person}{Jun Sun}, \bibinfo{person}{Yun Lin}, {and} \bibinfo{person}{Quang~Tran Minh}.} \bibinfo{year}{2020}\natexlab{}.
\newblock \showarticletitle{sFuzz: An Efficient Adaptive Fuzzer for Solidity Smart Contracts}. In \bibinfo{booktitle}{\emph{Proceedings of the ACM/IEEE 42nd International Conference on Software Engineering}}.
\newblock


\bibitem[Nikoli{\'c} et~al\mbox{.}(2018)]%
        {maian}
\bibfield{author}{\bibinfo{person}{Ivica Nikoli{\'c}}, \bibinfo{person}{Aashish Kolluri}, \bibinfo{person}{Ilya Sergey}, \bibinfo{person}{Prateek Saxena}, {and} \bibinfo{person}{Aquinas Hobor}.} \bibinfo{year}{2018}\natexlab{}.
\newblock \showarticletitle{Finding the Greedy, Prodigal, and Suicidal Contracts at Scale}. In \bibinfo{booktitle}{\emph{Proceedings of the 34th Annual Computer Security Applications Conference}}.
\newblock


\bibitem[Openchain(2023)]%
        {openchainapi}
\bibfield{author}{\bibinfo{person}{Openchain}.} \bibinfo{year}{2023}\natexlab{}.
\newblock \bibinfo{title}{Transaction Tracer}.
\newblock
\newblock
\urldef\tempurl%
\url{https://openchain.xyz/trace}
\showURL{%
\tempurl}


\bibitem[OpenSea(2018)]%
        {opensearegistry}
\bibfield{author}{\bibinfo{person}{OpenSea}.} \bibinfo{year}{2018}\natexlab{}.
\newblock \bibinfo{title}{WyvernProxyRegistry}.
\newblock
\newblock
\urldef\tempurl%
\url{https://etherscan.io/address/0xa5409ec958c83c3f309868babaca7c86dcb077c1}
\showURL{%
\tempurl}


\bibitem[OpenSea(2023)]%
        {opensea}
\bibfield{author}{\bibinfo{person}{OpenSea}.} \bibinfo{year}{2023}\natexlab{}.
\newblock \bibinfo{title}{OpenSea}.
\newblock
\newblock
\urldef\tempurl%
\url{https://opensea.io/}
\showURL{%
\tempurl}


\bibitem[OpenZeppelin(2021)]%
        {uups_vulnerability}
\bibfield{author}{\bibinfo{person}{OpenZeppelin}.} \bibinfo{year}{2021}\natexlab{}.
\newblock \bibinfo{title}{UUPSUpgradeable Vulnerability in OpenZeppelin Contracts}.
\newblock
\newblock
\urldef\tempurl%
\url{https://github.com/OpenZeppelin/openzeppelin-contracts-upgradeable/security/advisories/GHSA-q4h9-46xg-m3x9}
\showURL{%
\tempurl}


\bibitem[OpenZeppelin(2023a)]%
        {openzeppelin_github}
\bibfield{author}{\bibinfo{person}{OpenZeppelin}.} \bibinfo{year}{2023}\natexlab{a}.
\newblock \bibinfo{title}{Openzeppelin Contracts Upgradeable}.
\newblock
\newblock
\urldef\tempurl%
\url{https://github.com/OpenZeppelin/openzeppelin-contracts-upgradeable}
\showURL{%
\tempurl}


\bibitem[OpenZeppelin(2023b)]%
        {ownableupgradeable_sol}
\bibfield{author}{\bibinfo{person}{OpenZeppelin}.} \bibinfo{year}{2023}\natexlab{b}.
\newblock \bibinfo{title}{OwnableUpgradeable}.
\newblock
\newblock
\urldef\tempurl%
\url{https://github.com/OpenZeppelin/openzeppelin-contracts-upgradeable/blob/master/contracts/access/OwnableUpgradeable.sol}
\showURL{%
\tempurl}


\bibitem[OpenZeppelin(2023c)]%
        {uupsupgrabeable_sol}
\bibfield{author}{\bibinfo{person}{OpenZeppelin}.} \bibinfo{year}{2023}\natexlab{c}.
\newblock \bibinfo{title}{UUPSUpgradeable}.
\newblock
\newblock
\urldef\tempurl%
\url{https://github.com/OpenZeppelin/openzeppelin-contracts-upgradeable/blob/master/contracts/proxy/utils/UUPSUpgradeable.sol}
\showURL{%
\tempurl}


\bibitem[Ortner and Eskandari(2023)]%
        {smart_contract_sanctuary}
\bibfield{author}{\bibinfo{person}{Martin Ortner} {and} \bibinfo{person}{Shayan Eskandari}.} \bibinfo{year}{2023}\natexlab{}.
\newblock \showarticletitle{Smart Contract Sanctuary}.
\newblock  (\bibinfo{year}{2023}).
\newblock
\urldef\tempurl%
\url{https://github.com/tintinweb/smart-contract-sanctuary}
\showURL{%
\tempurl}


\bibitem[OwnableDelegateProxy(2018)]%
        {ownabledelegateproxy}
\bibfield{author}{\bibinfo{person}{OwnableDelegateProxy}.} \bibinfo{year}{2018}\natexlab{}.
\newblock \bibinfo{title}{OwnableDelegateProxy}.
\newblock
\newblock
\urldef\tempurl%
\url{https://etherscan.io/address/0x9b9c9daea6d5bf242fb1885b57d99a5a74433176}
\showURL{%
\tempurl}


\bibitem[Palladino(2020)]%
        {Stateofupgrade}
\bibfield{author}{\bibinfo{person}{Santiago Palladino}.} \bibinfo{year}{2020}\natexlab{}.
\newblock \bibinfo{title}{The State of Smart Contract Upgrades}.
\newblock
\newblock
\urldef\tempurl%
\url{https://blog.openzeppelin.com/the-state-of-smart-contract-upgrades}
\showURL{%
\tempurl}


\bibitem[Permenev et~al\mbox{.}(2020)]%
        {verx}
\bibfield{author}{\bibinfo{person}{Anton Permenev}, \bibinfo{person}{Dimitar Dimitrov}, \bibinfo{person}{Petar Tsankov}, \bibinfo{person}{Dana Drachsler-Cohen}, {and} \bibinfo{person}{Martin Vechev}.} \bibinfo{year}{2020}\natexlab{}.
\newblock \showarticletitle{Verx: Safety Verification of Smart Contracts}. In \bibinfo{booktitle}{\emph{2020 IEEE Symposium on Security and Privacy}}.
\newblock


\bibitem[Qin et~al\mbox{.}(2022)]%
        {BEV}
\bibfield{author}{\bibinfo{person}{Kaihua Qin}, \bibinfo{person}{Liyi Zhou}, {and} \bibinfo{person}{Arthur Gervais}.} \bibinfo{year}{2022}\natexlab{}.
\newblock \showarticletitle{Quantifying Blockchain Extractable Value: How Dark Is the Forest?}. In \bibinfo{booktitle}{\emph{2022 IEEE Symposium on Security and Privacy}}.
\newblock


\bibitem[Rodler et~al\mbox{.}(2019)]%
        {sereum}
\bibfield{author}{\bibinfo{person}{Michael Rodler}, \bibinfo{person}{Wenting Li}, \bibinfo{person}{Ghassan~O Karame}, {and} \bibinfo{person}{Lucas Davi}.} \bibinfo{year}{2019}\natexlab{}.
\newblock \showarticletitle{Sereum: Protecting Existing Smart Contracts Against Re-Entrancy Attacks}. In \bibinfo{booktitle}{\emph{26th Annual Network and Distributed System Security Symposium}}.
\newblock


\bibitem[Rodler et~al\mbox{.}(2021)]%
        {EVMPatch}
\bibfield{author}{\bibinfo{person}{Michael Rodler}, \bibinfo{person}{Wenting Li}, \bibinfo{person}{Ghassan~O Karame}, {and} \bibinfo{person}{Lucas Davi}.} \bibinfo{year}{2021}\natexlab{}.
\newblock \showarticletitle{EVMPatch: Timely and Automated Patching of Ethereum Smart Contracts}. In \bibinfo{booktitle}{\emph{30th USENIX Security Symposium}}.
\newblock


\bibitem[Shi et~al\mbox{.}(2022)]%
        {Classification}
\bibfield{author}{\bibinfo{person}{Chaochen Shi}, \bibinfo{person}{Yong Xiang}, \bibinfo{person}{Jiangshan Yu}, \bibinfo{person}{Longxiang Gao}, \bibinfo{person}{Keshav Sood}, {and} \bibinfo{person}{Robin Ram~Mohan Doss}.} \bibinfo{year}{2022}\natexlab{}.
\newblock \showarticletitle{A Bytecode-Based Approach for Smart Contract Classification}. In \bibinfo{booktitle}{\emph{2022 IEEE International Conference on Software Analysis, Evolution and Reengineering}}.
\newblock


\bibitem[So et~al\mbox{.}(2020)]%
        {verismart}
\bibfield{author}{\bibinfo{person}{Sunbeom So}, \bibinfo{person}{Myungho Lee}, \bibinfo{person}{Jisu Park}, \bibinfo{person}{Heejo Lee}, {and} \bibinfo{person}{Hakjoo Oh}.} \bibinfo{year}{2020}\natexlab{}.
\newblock \showarticletitle{VeriSmart: A Highly Precise Safety Verifier for Ethereum Smart Contracts}. In \bibinfo{booktitle}{\emph{2020 IEEE Symposium on Security and Privacy}}.
\newblock


\bibitem[Solidity(2023)]%
        {solidity}
\bibfield{author}{\bibinfo{person}{Solidity}.} \bibinfo{year}{2023}\natexlab{}.
\newblock \bibinfo{title}{Solidity Programming Language}.
\newblock
\newblock
\urldef\tempurl%
\url{https://soliditylang.org/}
\showURL{%
\tempurl}


\bibitem[Stephens et~al\mbox{.}(2021)]%
        {SmartPulse}
\bibfield{author}{\bibinfo{person}{Jon Stephens}, \bibinfo{person}{Kostas Ferles}, \bibinfo{person}{Benjamin Mariano}, \bibinfo{person}{Shuvendu Lahiri}, {and} \bibinfo{person}{Isil Dillig}.} \bibinfo{year}{2021}\natexlab{}.
\newblock \showarticletitle{SmartPulse: Automated Checking of Temporal Properties in Smart Contracts}. In \bibinfo{booktitle}{\emph{2021 IEEE Symposium on Security and Privacy}}.
\newblock


\bibitem[Tikhomirov et~al\mbox{.}(2018)]%
        {smartchecker}
\bibfield{author}{\bibinfo{person}{Sergei Tikhomirov}, \bibinfo{person}{Ekaterina Voskresenskaya}, \bibinfo{person}{Ivan Ivanitskiy}, \bibinfo{person}{Ramil Takhaviev}, \bibinfo{person}{Evgeny Marchenko}, {and} \bibinfo{person}{Yaroslav Alexandrov}.} \bibinfo{year}{2018}\natexlab{}.
\newblock \showarticletitle{Smartcheck: Static Analysis of Ethereum Smart Contracts}. In \bibinfo{booktitle}{\emph{Proceedings of the 1st International Workshop on Emerging Trends in Software Engineering for Blockchain}}.
\newblock


\bibitem[Torres et~al\mbox{.}(2021)]%
        {confuzzius}
\bibfield{author}{\bibinfo{person}{Christof~Ferreira Torres}, \bibinfo{person}{Antonio~Ken Iannillo}, \bibinfo{person}{Arthur Gervais}, {and} \bibinfo{person}{Radu State}.} \bibinfo{year}{2021}\natexlab{}.
\newblock \showarticletitle{Confuzzius: A Data Dependency-Aware Hybrid Fuzzer for Smart Contracts}. In \bibinfo{booktitle}{\emph{2021 IEEE European Symposium on Security and Privacy}}.
\newblock


\bibitem[Torres et~al\mbox{.}(2018)]%
        {osiris}
\bibfield{author}{\bibinfo{person}{Christof~Ferreira Torres}, \bibinfo{person}{Julian Sch{\"u}tte}, {and} \bibinfo{person}{Radu State}.} \bibinfo{year}{2018}\natexlab{}.
\newblock \showarticletitle{Osiris: Hunting for Integer Bugs in Ethereum Smart Contracts}. In \bibinfo{booktitle}{\emph{Proceedings of the 34th Annual Computer Security Applications Conference}}.
\newblock


\bibitem[Truffle(2018)]%
        {truffle-code-utils}
\bibfield{author}{\bibinfo{person}{Truffle}.} \bibinfo{year}{2018}\natexlab{}.
\newblock \bibinfo{title}{Truffle-code-utils}.
\newblock
\newblock
\urldef\tempurl%
\url{https://www.npmjs.com/package/truffle-code-utils}
\showURL{%
\tempurl}


\bibitem[Uniswap(2020)]%
        {token_list}
\bibfield{author}{\bibinfo{person}{Uniswap}.} \bibinfo{year}{2020}\natexlab{}.
\newblock \bibinfo{title}{Introducing Token Lists}.
\newblock
\newblock
\urldef\tempurl%
\url{https://blog.uniswap.org/token-lists}
\showURL{%
\tempurl}


\bibitem[Uniswap(2023)]%
        {uniswap}
\bibfield{author}{\bibinfo{person}{Uniswap}.} \bibinfo{year}{2023}\natexlab{}.
\newblock \bibinfo{title}{Uniswap}.
\newblock
\newblock
\urldef\tempurl%
\url{https://uniswap.org/}
\showURL{%
\tempurl}


\bibitem[Vogelsteller(2015)]%
        {eip_20}
\bibfield{author}{\bibinfo{person}{Fabian Vogelsteller}.} \bibinfo{year}{2015}\natexlab{}.
\newblock \bibinfo{title}{Token Standard}.
\newblock
\newblock
\urldef\tempurl%
\url{https://eips.ethereum.org/EIPS/eip-20}
\showURL{%
\tempurl}


\bibitem[Wang et~al\mbox{.}(2019)]%
        {NPChecker}
\bibfield{author}{\bibinfo{person}{Shuai Wang}, \bibinfo{person}{Chengyu Zhang}, {and} \bibinfo{person}{Zhendong Su}.} \bibinfo{year}{2019}\natexlab{}.
\newblock \showarticletitle{Detecting Nondeterministic Payment Bugs in Ethereum Smart Contracts}.
\newblock \bibinfo{journal}{\emph{Proceedings of the ACM on Programming Languages}}.
\newblock


\bibitem[Web3(2023)]%
        {getcode}
\bibfield{author}{\bibinfo{person}{Web3}.} \bibinfo{year}{2023}\natexlab{}.
\newblock \bibinfo{title}{getCode}.
\newblock
\newblock
\urldef\tempurl%
\url{https://web3js.readthedocs.io/en/v1.2.11/web3-eth.html#getcode}
\showURL{%
\tempurl}


\bibitem[Xue et~al\mbox{.}(2020)]%
        {Clairvoyance}
\bibfield{author}{\bibinfo{person}{Yinxing Xue}, \bibinfo{person}{Mingliang Ma}, \bibinfo{person}{Yun Lin}, \bibinfo{person}{Yulei Sui}, \bibinfo{person}{Jiaming Ye}, {and} \bibinfo{person}{Tianyong Peng}.} \bibinfo{year}{2020}\natexlab{}.
\newblock \showarticletitle{Cross-Contract Static Analysis for Detecting Practical Reentrancy Vulnerabilities in Smart Contracts}. In \bibinfo{booktitle}{\emph{Proceedings of the 35th IEEE/ACM International Conference on Automated Software Engineering}}.
\newblock


\bibitem[Xue et~al\mbox{.}(2022)]%
        {xfuzzer}
\bibfield{author}{\bibinfo{person}{Yinxing Xue}, \bibinfo{person}{Jiaming Ye}, \bibinfo{person}{Wei Zhang}, \bibinfo{person}{Jun Sun}, \bibinfo{person}{Lei Ma}, \bibinfo{person}{Haijun Wang}, {and} \bibinfo{person}{Jianjun Zhao}.} \bibinfo{year}{2022}\natexlab{}.
\newblock \showarticletitle{xFuzz: Machine Learning Guided Cross-Contract Fuzzing}.
\newblock \bibinfo{journal}{\emph{IEEE Transactions on Dependable and Secure Computing}} (\bibinfo{year}{2022}).
\newblock


\bibitem[YAcademy(2022)]%
        {security_guide}
\bibfield{author}{\bibinfo{person}{YAcademy}.} \bibinfo{year}{2022}\natexlab{}.
\newblock \bibinfo{title}{Security Guide to Proxies}.
\newblock
\newblock
\urldef\tempurl%
\url{https://proxies.yacademy.dev/pages/security-guide/}
\showURL{%
\tempurl}


\bibitem[Zhang et~al\mbox{.}(2020)]%
        {smartfield}
\bibfield{author}{\bibinfo{person}{Yuyao Zhang}, \bibinfo{person}{Siqi Ma}, \bibinfo{person}{Juanru Li}, \bibinfo{person}{Kailai Li}, \bibinfo{person}{Surya Nepal}, {and} \bibinfo{person}{Dawu Gu}.} \bibinfo{year}{2020}\natexlab{}.
\newblock \showarticletitle{Smartshield: Automatic Smart Contract Protection Made Easy}. In \bibinfo{booktitle}{\emph{2020 IEEE 27th International Conference on Software Analysis, Evolution and Reengineering}}.
\newblock


\end{thebibliography}

\newpage

\appendix

\section{APPENDIX}

\begin{table}
\footnotesize
  \caption{Opcodes and Function Information Extracted From Bytecode.}
  \vspace{-3mm}
  \hspace*{-2mm}
  \label{ta:opcodes}
  \centering
  \begin{threeparttable}

    \begin{tabular}{lll} 

 \toprule

     \textbf{Infor.} & \textbf{Abbr.} & \textbf{Desc.}  \\ 

     \midrule[0.5pt]

     \cellcolor[HTML]{DFDFDF}CALL & \cellcolor[HTML]{DFDFDF}CALL & \cellcolor[HTML]{DFDFDF}Call a method in another contract.  \\

     \multirow{2}{*}{DELEGATECALL} &  \multirow{2}{*}{DCALL} & Call a method in another contract \\ 
     &  & using the storage of current contract.  \\

    \cellcolor[HTML]{DFDFDF} &\cellcolor[HTML]{DFDFDF} & \cellcolor[HTML]{DFDFDF}Call a method in another contract \\
      \multirow{-2}{*}{\cellcolor[HTML]{DFDFDF}STATICCALL} & \multirow{-2}{*}{\cellcolor[HTML]{DFDFDF}SCALL} & \cellcolor[HTML]{DFDFDF}without state changes. \\

     SELFDESTRUCT & SDES &  Destroy the contract.  \\

     \cellcolor[HTML]{DFDFDF} &\cellcolor[HTML]{DFDFDF} & \cellcolor[HTML]{DFDFDF}Fallback function will be called when a non- \\
      \multirow{-2}{*}{\cellcolor[HTML]{DFDFDF}Fallback} & \multirow{-2}{*}{\cellcolor[HTML]{DFDFDF}FBK} & \cellcolor[HTML]{DFDFDF}existent function is called on the current contract. \\

      Functions & \{Func\} & A set of functions belong to current contract.   \\ 

     \cellcolor[HTML]{DFDFDF} &\cellcolor[HTML]{DFDFDF} & \cellcolor[HTML]{DFDFDF}A set of functions that current contract calls \\
      \multirow{-2}{*}{\cellcolor[HTML]{DFDFDF}Other Functions} & \multirow{-2}{*}{\cellcolor[HTML]{DFDFDF}\{OFunc\}} & \cellcolor[HTML]{DFDFDF}another contract. \\
      
     \bottomrule
     
    \end{tabular}
  \end{threeparttable}
  \end{table}

\begin{table}
\footnotesize
  \caption{The Rules Specified for Each Pattern. }
  \vspace{-3mm}
  \hspace*{-2mm}
  \label{ta:rules}
  \centering
  \begin{threeparttable}

    \begin{tabular}{l@{}l} 
 \toprule

      \textbf{Pattern} & \textbf{~Rules~} \\ \midrule[0.5pt]
     
      \cellcolor[HTML]{DFDFDF}Proxy Pattern & \cellcolor[HTML]{DFDFDF}$~$ $Upg \in (\{Func\}_{proxy} \lor \{Func\}_{logic}) \land  DCALL \land FBK $  \\  

      Strategy Pattern & $~$ $Upg\in \{Func\}_{main} \land (CALL \lor SCALL) \land \{OFunc\}_{main} $  \\ 

      \cellcolor[HTML]{DFDFDF}Data Separation & \cellcolor[HTML]{DFDFDF}$~$ $Upg \in \{Func\}_{data} \lor Rule_{strategy}$ \\ 

      Mix Pattern & $~$ $Rule_{proxy} \land Rule_{strategy}$ \\ 
    
     \cellcolor[HTML]{DFDFDF}Metamorphic Contract & \cellcolor[HTML]{DFDFDF}$~$ $SDES \land CREATE2 \in Call Trace_{tx\_create}$ \\ 
     
      Contract Migration & $~$ $Addr_{old}  \land  Announcement \land  Addr_{new}$ \\ 

     \bottomrule
    \end{tabular}
  \end{threeparttable}
  \end{table}

{\noindent{\textbf{\large I. USCDetector Details}}\hfill}
\label{app1}

{\noindent \textbf{\underline{Notations.}}}
\newline

\begin{table}[h]
\small
\vspace{-3mm}
  \centering
  \begin{threeparttable}

    \begin{tabular}{lp{5.6cm}} 
    $Upg$ &: Upgrade Function.  \\ 
    $\{Func\}_{proxy}$ &: A set of functions belong to proxy.  \\ 
    $\{OFunc\}_{main}$ &: A set of functions that the main contract calls the logic contract.   \\ 
    $Rule_{strategy}$ &: The rule for detecting strategy pattern.  \\ 
    $Call Trace_{tx\_create}$ &: The call traces of the transaction that creates the contract.  \\ 
    $Addr_{old}$ &: The address of the old contract.  \\
    $Announcement$ &: The upgrade announcement obtained from Etherscan.  
     
    \end{tabular}
  \end{threeparttable}
  \vspace{-2mm}
  \end{table}

{\noindent \textbf{\underline{Rules.}}} 
\newline

\textbf{Proxy Pattern Rule} requires (1) \textit{DELEGATECALL}, (2)\textit{fallback} function, and (3) the upgrade function must exist either in the proxy or logic contract.

\textbf{Strategy Pattern Rule } requires three elements to exist in the main contract at the same time. These elements are the upgrade function, \textit{CALL} or \textit{STATICCALL} (or both of them), and the functions that the main calls the logic contract.

\textbf{Data Separation Rule} requires the upgrade function to exist in the data contract. If the data contract calls functions of other contracts, it then has the same features as the strategy pattern. 

\textbf{Mix Pattern Rule } requires that both the proxy's rule and strategy's rule must be satisfied.

\textbf{Metamorphic Contract Rule} requires \textit{SELFDESTRUCT} to exist in the contract, while the call trace of the transaction that creates the contract must contain \textit{CREATE2}.

\textbf{Contract Migration Rule} requires the address of the old contract, and announcement, and the address of the new contract must exist in the records we obtain from Etherscan at the same time.

\clearpage

{\noindent{\textbf{\large II. False Negative Evaluation Using Dataset from~\cite{ProxyHunting}}}\hfill}
\label{app4}

We evaluate USCDetector using a subset of the dataset from Proxy Hunting ~\cite{ProxyHunting}, including 775 smart contracts. 
The TP, FP, TN, and FN are 673, 1, 94, and 7, respectively. 
Particularly, for 2 out of 7 FNs, our dataset missed the upgrading functions. For the rest 5, the logic contract address is hardcoded in the source code of the proxy contract. 
Our approach cannot obtain it from the RPC and cause the FN. 
In particular, there are two reasons to only use a subset of their dataset: 
(1) As mentioned in their paper, Proxy Hunting cannot detect USCs if the source code of the logic contract is not available, while our approach relying on bytecode can. 
(2) We find that inconsistency between source code and bytecode exists in some cases. Basically, the source code provided by the developer at an address does not exactly match the bytecode deployed at that address. 
The reason could be that the developer attempts to provide all relevant source code to verify a smart contract, while the deployed contract is only part of it. 
For example, the contract \textit{CoreLibrary}~\cite{corelibrary} is the deployed contract and does not have any upgrading function. 
However, the source code provided by the developer at this address also includes the code for   \textit{AdminUpgradeabilityProxy} (which is a proxy-based USC) and its dependent contracts. 
We thus excluded these contracts in our evaluation.

{\noindent{\textbf{\large III. Additional Listings of Vulnerable Examples}}\hfill}
\label{app2}

List~\ref{list1} presents a simplified version of a Mix Pattern contract that has no restrictive check on contract admin, derived from a real-world DApp LANDProxy~\cite{decentraland}. 
There is no admin check on the \textit{upgrade} function, and thus anyone can overwrite the logic address (Line 3).

\begin{lstlisting}[caption=A simplified version of
 LANDProxy,label=list1,
morekeywords={contract, function, public, require},
emph={bytes, payable},
    emphstyle=\color{teal}, 
    ]
contract Proxy {
    function upgrade(IApplication newContract, bytes data) public {
        currentContract = newContract;
        newContract.initialize(data);
    }
    function () payable public {
        require(currentContract != 0);
        delegatedFwd(currentContract, msg.data);
    } ...
}
    
\end{lstlisting}

Listing~\ref{list2} presents a case I example. It lists the decompiled code of a simplified version of a (unverified) vulnerable UUPS-based logic contract. 
There is a function \texttt{initialize} (Line 8) that can be called directly. 
This function can declare the owner of the contract (Line 9). 
Since this contract has no state, its ownership can be obtained by calling the \texttt{initialize} function. 
Attackers can further call the \texttt{destruct} (Line 4) function to destroy this contract, disabling its proxy contract and withdrawing all the ETHs (Line 7).

\begin{lstlisting}[caption=Decompiled code of a Logic Contract that Contains SELFDESTRUCT,label=list2,
morekeywords={def, is, selfdestruct, if, else, revert},
emph={stor0, stor1, owner, },
    emphstyle=\color{teal}, 
    ]
def storage:
  owner is address at storage 151
    ...
def destruct(address to):
  if owner != caller:
      revert 'not the owner'
  selfdestruct(to)
def initialize(): 
  owner = caller
\end{lstlisting}

Listing~\ref{list3} presents a case II example: a simplified version of a UUPS-based logic contract. 
It does not contain {\itshape SELFDESTRUCT} opcode, but can still be destroyed through the function \texttt{upgradeToAndCall}. 
The upgrade mechanism is implemented correctly for the proxy: only the owner can perform upgrades (Line 10). 
However, the \texttt{upgradeToAndCall} function can be directly called by the logic contract owner.
An attacker can call the function \texttt{initialize} (Line 3) to take ownership of the logic contract, and further destroy it by calling the function \texttt{upgradeToAndCall} (Line 10). 

Particularly, the \texttt{\_upgradeToAndCallSecure} function includes a rollback test to validate that the new logic address also has an upgrade function (Line 29). 
However, this test can be bypassed by performing twice upgrades: first by resetting the {\itshape rollbackTesting} value (Line 28) and then by calling \texttt{\_upgradeToAndCallSecure} function to a function containing {\itshape SELFDESTRUCT} in the upgrading logic contract (Line 26). 
This would destroy the logic contract and cause the proxy's {\itshape DELEGATECALL} to point to a self-destructed logic contract.

\begin{lstlisting}[caption=Decompiled code of a Logic Contract that Function Setter Can Be Called Directly, label=list3,
morekeywords={contract, abstract, function, virtual, string, public, external, is, true, internal},
emph={payable, address, memory, bool, bytes},
    emphstyle=\color{teal}, 
    ]
contract Token is ..., UUPSUpgradeable{
    ...
    function initialize(string memory _name, string memory _symbol) initializer public {
        ...
     }
     ...
}
abstract contract UUPSUpgradeable is ERC1967UpgradeUpgradeable {
    ...
    function upgradeToAndCall(address newImplementation, bytes memory data) external payable virtual {
        _authorizeUpgrade(newImplementation);
        _upgradeToAndCallSecure(newImplementation, data, true);
    }
    ...
}
abstract contract ERC1967UpgradeUpgradeable {
    ...
    function _upgradeToAndCallSecure(
        address newImplementation,
        bytes memory data,
        bool forceCall
    ) internal {
        ...
        _setImplementation(newImplementation);
        if (data.length > 0 || forceCall) {
            _functionDelegateCall(newImplementation, data);
        }
        BooleanSlot storage rollbackTesting = getBooleanSlot(_ROLLBACK_SLOT);
        if (!rollbackTesting.value) {
            ...
            _upgradeTo(newImplementation);
        }
    }
    ...
}
...
    
\end{lstlisting}

\newpage
{\noindent{\textbf{\large IV. Additional Figures of Vulnerable Examples}}\hfill}
\label{app5}

\begin{figure}[h]
    \vspace{-2mm}
\hspace{-3mm}
\begin{subfigure}{\linewidth}
    \includegraphics[scale=0.55]{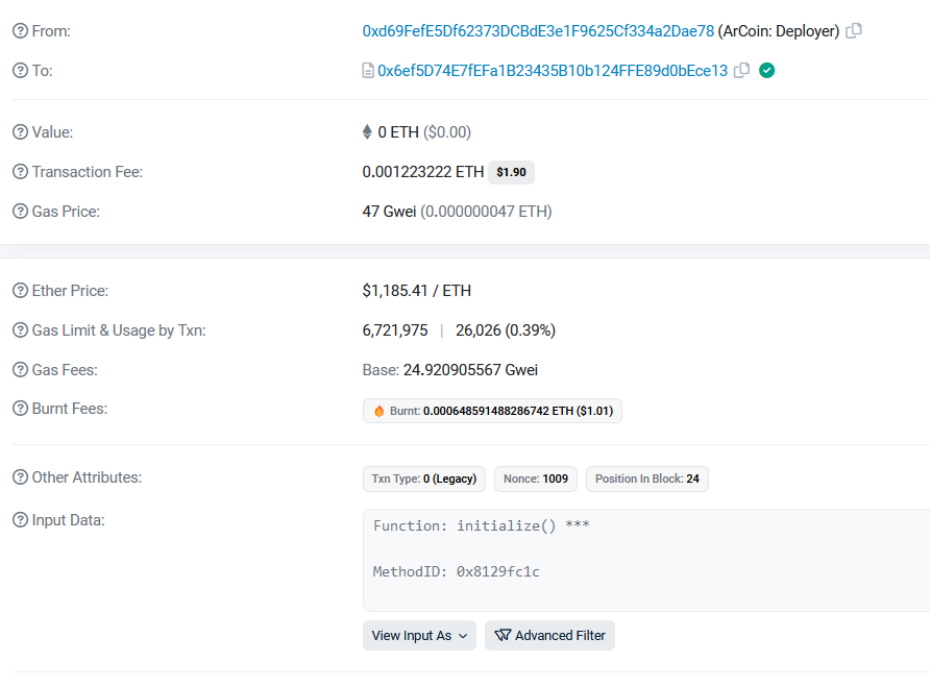}
    \vspace{-3mm}
    \caption{ A Transaction Send from Contract's Admin}
    \label{fig:example1}
\end{subfigure} 
\hspace{-3mm}
\begin{subfigure}{\linewidth}
    \includegraphics[scale=0.4]{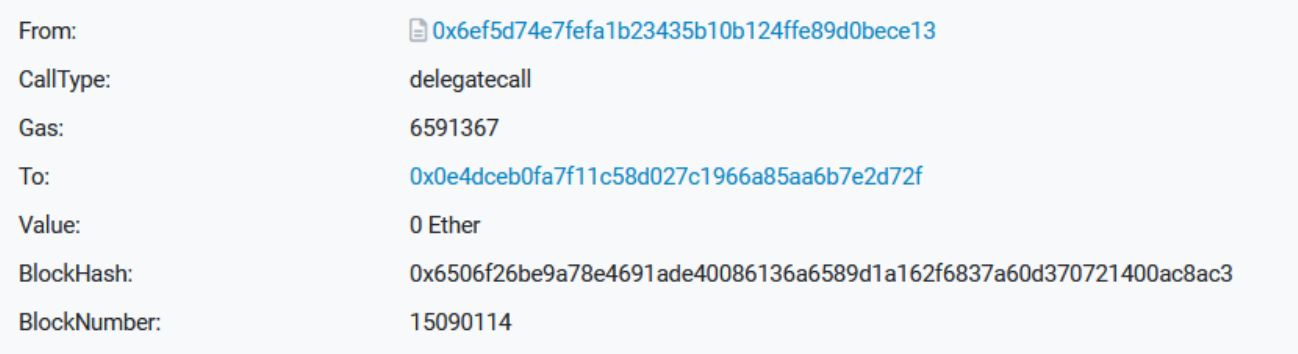}
    \vspace{-3mm}
    \caption{The Proxy Contract Delegate Call to an EOA }
    \label{fig:example2}
\end{subfigure}
\vspace{-4mm}
\caption{USC Patterns.}
\label{fig: eoa_example}
\vspace{-3mm}
\end{figure}

Figure~\ref{fig: eoa_example} presents a real-world example of proxy-based USC. Figure~\ref{fig:example1} shows a transaction from a proxy's admin to {\itshape initialize} the proxy contract. 
As the function {\itshape initialize } is supposed to exist in the logic contract, the proxy contract simply delegates the call to the logic contract (Figure~\ref{fig:example2}). 
However, the logic address is actually an EOA.
 Thus, the {\itshape initialize} was not executed and eventually has no state change.

\end{document}